\documentclass[preprint2]{aastex}

\newcommand{\eg}{{\rm e.g.\ }}
\newcommand{\ie}{{\rm i.e.\ }}

\newcommand{\etal}{{\rm et al.\ }}

\newcommand{\gsim}{\gtrsim}
\newcommand{\lsim}{\lesssim}

\newcommand{\Msol}{M_\odot}
\newcommand{\Lsol}{L_\odot}
\newcommand{\Zsol}{Z_\odot}

\newcommand{\ergs}{{\rm erg\,s^{-1}}}
\newcommand{\Hz}{{\rm Hz}}
\newcommand{\pc}{{\rm pc}}
\newcommand{\kpc}{{\rm kpc}}
\newcommand{\Mpc}{{\rm Mpc}}
\newcommand{\kms}{{\rm km\,s^{-1}}}
\renewcommand{\mag}{{\rm mag\,}}
\newcommand{\mum}{\micron}
\newcommand{\Myr} {{\rm Myr}}

\newcommand{\yr }{{\rm yr}}

\newcommand{\Vhot} { V_{\rm hot}}

\newcommand{\Vdisk} { V_{\rm disk}}
\newcommand{\Vbulge} { V_{\rm bulge}}

\newcommand{\alphahot}{\alpha_{\rm hot}}
\newcommand{\alphastar}{\alpha_{\star}}

\newcommand{\fellip}{f_{\rm ellip}}
\newcommand{\Mdisk}{M_{\rm disk}}
\newcommand{\Mbulge}{M_{\rm bulge}}
\newcommand{\rdisk}{r_{\rm disk}}
\newcommand{\rbulge}{r_{\rm bulge}}
\newcommand{\taudisk} { \tau_{\rm disk}}
\newcommand{\taubulge} { \tau_{\rm bulge}}
\newcommand{\epsdisk}{\epsilon_{\rm \star disk}}
\newcommand{\epsburst}{\epsilon_{\rm \star burst}}
\newcommand{\taustdisk}{\tau_{\rm \star disk}}
\newcommand{\taustburst}{\tau_{\rm \star burst}}
\newcommand{\betadisk} {\beta_{\rm disk}}
\newcommand{\betaburst} {\beta_{\rm burst}}
\newcommand{\rburst} {r_{\rm burst}}
\newcommand{\Mburst} {M_{\rm burst}}
\newcommand{\Mcold}{M_{\rm cold}}
\newcommand{\Zcold}{Z_{\rm cold}}
\newcommand{\tesc} {t_{\rm esc}}
\newcommand{\Mdust}{M_{\rm dust}}
\newcommand{\Mstar}{M_{\rm star}}
\newcommand{\Zstar}{Z_{\star}}
\newcommand{\Zhot}{Z_{\rm hot}}
\shorttitle{Galaxy formation models in the infrared}
\shortauthors{Granato et al. }

\begin{document}


\title{The infrared side of galaxy formation. \\
I. The local universe in the semi-analytical framework.}


\author{G.L. Granato\altaffilmark{1,4,5}, C.G. Lacey\altaffilmark{2,3,4},
L. Silva\altaffilmark{4,1}, A. Bressan\altaffilmark{1},
C.M. Baugh\altaffilmark{3}, S. Cole\altaffilmark{3},
C.S.~Frenk\altaffilmark{3}}

\altaffiltext{1}{Osservatorio Astronomico di Padova, Vicolo
dell'Osservatorio, 5, I-35122 Padova, Italy}

\altaffiltext{2}{Theoretical Astrophysics Center, Juliane Maries Vej
30, 2100 Copenhagen \O, Denmark} 

\altaffiltext{3}{Physics Department, Durham University, South Road,
Durham DH1 3LE, UK}

\altaffiltext{4}{SISSA, Via Beirut 2-4, I-34014 Trieste, Italy}

\altaffiltext{5}{granato@pd.astro.it}


\begin{abstract}

We present a new evolutionary model for predicting the far-UV to sub-mm
properties of the galaxy population. This combines a semi-analytic galaxy
formation model based on hierarchical clustering
\citep[GALFORM,][]{cole99} with
a spectro-photometric code which includes dust reprocessing
\citep[GRASIL,][]{silva98}. The former provides the star formation and metal
enrichment histories, together with the gas mass and various geometrical
parameters, for a representative sample of galaxies formed in different density
environments. These quantities, together with a few other assumptions
concerning the spatial distribution of dust and its optical properties, allow
us to model the spectral energy distributions (SEDs) of galaxies, taking into
account stellar emission and also dust extinction (absorption plus scattering)
and re-emission. In the spectro-photometric code dust is considered only in the
disk, but the general radiation field is contributed by both the disk and the
bulge components with their own distinct age and metallicity distributions. Two
phases are considered for the dust: molecular cloud complexes, where stars are
assumed to be born, and the diffuse interstellar medium. The model includes
both galaxies forming stars quiescently in disks, and starbursts triggered by
galaxy mergers. We test our models against the observed spectro-photometric
properties of galaxies in the local Universe, assuming a CDM cosmology with
$\Omega_0=0.3$ and $\Lambda_0=0.7$. The models reproduce fairly well
the SEDs of 
normal spirals and starbursts from the far-UV to the sub-mm, and their internal
extinction properties. The starbursts follow the observed relationship between
the FIR to UV luminosity ratio and the slope of the UV continuum. They also
reproduce the observed starburst attenuation law \citep{calz99a}.  This result
is remarkable, because we use a dust mixture which reproduces the Milky Way
extinction law. It suggests that the observed attenuation law is closely
related to the geometry of the stars and dust. We compute galaxy luminosity
functions over a wide range of wavelengths, which turn out to be in good
agreement with observational data in the UV (2000\AA), in the B and K bands,
and in the IR ($12-100\mum$). Finally, we investigate the reliability of some
star formation indicators which are based on the properties of the continuum
SEDs of galaxies. The UV continuum turns out to be a poor star formation
indicator for our models, whilst the infrared luminosity is much more reliable.

\end{abstract}

\keywords{Galaxies: evolution;
Galaxies: formation; Galaxies: fundamental parameters; Galaxies: interactions;
Galaxies: starburst; Infrared: galaxies; Ultraviolet: galaxies}

\section{Introduction}
\label{sec:intro}

In recent years, our understanding of galaxy formation and evolution has
advanced very rapidly, as a result of both observations and theory. On the
observational side, new instruments have allowed the direct study of galaxy
populations at different wavelengths out to $z\lsim 5$. By combining
observations in the UV, optical, IR and sub-mm, we can now start to reconstruct
the history of star formation in galaxies over the epochs when the bulk of the
stars have formed \citep[\eg][]{madau96,steidel99,hughes98}. On the theoretical
side, models based on the paradigm of structure formation through hierarchical
clustering (which has successfully confronted a wide range of observations on
large scale structure and microwave background anisotropies) have now been
developed to the point where they can make definite predictions for the
observable properties of galaxies (luminosities, colours, sizes, morphologies
etc) at all redshifts, starting from an assumed initial spectrum of density
fluctuations. The key technique for making these predictions has been that of
{\em semi-analytical modelling} \citep{wf91,lacey91,kauff93,cole94,sp99}. In
this technique, one applies simplified analytical descriptions of the main
physical processes of gas cooling and collapse, star formation, feedback
effects from supernovae, galaxy merging etc, with the backbone being a Monte
Carlo description of the process of formation and merging of dark matter halos
through hierarchical clustering. The predicted star formation histories are
then combined with detailed stellar population models to calculate galaxy
luminosities at different wavelengths. Conversely, direct numerical
simulations have been enormously successful in studying the evolution of
structure in the dark matter on a huge range of scales
\citep[\eg][]{jenkins98}, but currently do not have sufficient spatial
resolution to simultaneously follow all the processes involved in galaxy
formation.

The semi-analytical models have been successful in predicting and/or
explaining a large range of galaxy properties, both at low and high
redshift, for instance, luminosity functions and colours in different
optical and near-IR bands \citep{lacey93,kauff93,cole94}, the mixture
of galaxy morphologies and the evolution of elliptical galaxies
\citep{kauff93,baugh96,kauff96}, the properties of Lyman-break
galaxies at high redshift \citep{baugh98,governato98}, the sizes and
circular velocities of galaxies \citep{cole99}, and galaxy clustering
evolution and the nature of the clustering bias
\citep{kauff97,baugh99,diaferio99,benson99}. However, with very few
exceptions, these semi-analytical models have ignored both extinction
and emission by interstellar dust, and calculated only the direct
stellar emission in the UV, optical and near-IR. This has been partly
because the importance of dust was generally under-appreciated,
especially for high redshift galaxies, but also because of the lack of
physically realistic models for predicting dust effects.

This situation has now begun to change. On the one hand, there have been
several observational discoveries demonstrating the importance of dust effects
for building a complete picture of galaxy formation. (1) The discovery of a
cosmic far-IR/sub-mm background by the COBE satellite
\citep{puget96,guid97,dwek98,fixsen98,hauser98,schlegel98}, whose energy
density indicates that, as suggested already by \citet{wang91} and
\citet{franc94}, a large fraction of the energy radiated by stars over the
history of the universe has been reprocessed by dust.  (2) The discovery that
the population of star forming galaxies at $z\sim 2-4$ that have been detected
through their strong Lyman-break features are substantially extincted in the
rest-frame UV \citep{pettini98,steidel99}. (3) The discovery of a population of
sub-mm sources at high redshift ($z\gsim 1$) using SCUBA, whose luminosities,
if they are powered by star formation in dust-enshrouded galaxies, imply very
large star formation rates ($\sim 10^2 \Msol\yr^{-1}$), and a total star
formation density comparable to what is inferred from the UV luminosities of
the Lyman-break galaxies \citep{smail97,hughes98,lilly99}. (4) The ISO
detection of a population of strong IR sources; 15 $\mu$m ISOCAM
\citep[\eg][]{oliver97,elbaz99} and 175 $\mu$m ISOPHOT surveys
\citep[\eg][]{kawara98,puget99} indicate a population of actively star forming
galaxies at $0.4 \lsim z \lsim 1.3$, which boosts the cosmic star
formation density by a
factor $\sim 3$ with respect to that estimated in the optical from
the CFRS \citep{flores99}. For 
(1) and (3), there is the caveat that the contribution from dust-enshrouded
AGNs to the sub-mm counts and background is currently uncertain, but probably
the AGNs do not dominate \citep{gdf97,almaini99,madau99}. These discoveries
demonstrate that in order to understand the history of star formation in the
universe from observational data, one must have a unified picture that covers
all wavelengths from the far-UV to the sub-mm. The UV and the far-IR are
expecially important, since young stellar populations emit most of their
radiation in the rest-frame UV, but a significant fraction of this is dust
reprocessed into the rest-frame far-IR.

On the theoretical side, it is now possible for the first time to construct
true {\em ab initio} models in which the galaxy formation itself and stellar
emission and dust absorption and emission are calculated from first principles,
based on physical models, and avoiding observational parameterizations for
various key ingredients (\eg shape of the luminosity function, dependence of
dust temperature on galaxy properties). These new models, which provide a
unified treatment of emission from stars and dust, and predict the evolution of
galaxy luminosities from the far-UV to the mm, are the subject of this paper.

The effects of dust on galaxy luminosities at different wavelengths have been
included in some previous galaxy evolution models, at various levels of
sophistication, but mostly in the context of {\em backwards evolution} models,
where one tries to evolve back from observed galaxy properties at the present
day, in contrast to the {\em semi-analytical} models, where one evolves forward
from cosmological initial conditions. In backwards evolution models, one starts
from the observed luminosity functions of different types of galaxy at the
present day, assumes a different star formation history for each type, and
calculates the luminosity evolution for each type, to predict what the galaxy
population would have looked like in the past. \citet{grv87} were the first to
include {\em dust absorption} in a model of this type, based on a 1D slab model
for the star and dust distribution, and calculating the dust content
self-consistently from a chemical evolution model. The same treatment of dust
was later used in the semi-analytical galaxy formation models of
\citet{lacey93}. In both cases, the models were used to calculate galaxy
luminosities and number counts in the UV and optical. \citet{mazzei92} were the
first to try to model the evolution of stellar emission and dust emission
together in a consistent framework based on stellar population synthesis models
and a physical calculation of dust absorption. This model was then used by
\citet{franc94} to calculate galaxy evolution and number counts in bands from
the optical through to the far-IR, based on the backwards evolution approach.
However, these models still made a number of simplifying assumptions (\eg slab
geometry for disks), and set a number of present-day properties of galaxies
from observations (\eg the optical depth of galactic disks, and the intensity
of the radiation field heating the dust), rather than predicting
them. Simpler backwards evolution models, where the 
luminosity evolution is parameterized as a simple function of
redshift, have been considered by, \eg \citet{pearson96}.

Recently, dust absorption has been included in several different
semi-analytical models \citep{kauff99, sp99, cole99}. The first two 
calculate dust effects only for present-day galaxies, using a 1D slab model,
and taking the dust optical depth from observational measurements. On the other
hand, \citet{cole99} predict the dust optical depth and how it evolves, based
on chemical evolution and a prediction of disk sizes, and use the 3D disk+bulge
radiative transfer models of \citet{ferrara99} to calculation the dust
attenuation. The only previous semi-analytical model to calculate dust emission
as well as absorption is that of \citet{guider98}. However, that model also has
several limitations: the galaxy formation model does not include merging of
either dark halos or visible galaxies, and the fraction of star formation
occuring in bursts is simply an arbitrary function; dust absorption is again
modelled assuming a 1D slab geometry; and the dust temperature distribution is
not predicted. Instead, the dust emission spectrum is modelled as the sum of
several components, whose temperatures and relative strengths are chosen so as
to reproduce the observed correlations of IR colours with IR luminosity found
by IRAS. 

The present paper represents a major advance over this earlier work in terms of
scope, physical self-consistency and predictiveness. We combine the
semi-analytical galaxy formation model of \citet{cole99} with the stellar
population + dust model of \citet{silva98}. The galaxy formation model includes
formation of dark halos through merging, cooling and collapse of gas in halos
to form disks, star formation in disks regulated by energy input from
supernovae, merging of disk galaxies to form elliptical galaxies and bulges,
bursts of star formation triggered by these mergers, predictions of the radii
of disks and spheroids, and chemical enrichment of the stars and gas. The
stellar population + dust model includes a realistic 3D geometry, with a disk
and bulge, two phase dust in clouds and in the diffuse ISM, star formation in
the clouds, radiative transfer of starlight through the dust distribution, a
realistic dust grain model including PAHs and quantum heating of small grains,
and a direct prediction of the dust temperature distribution at each point in
the galaxy based on a calculation of dust heating and cooling. The output is
the luminosity and spectrum of the stellar populations attenuated by dust, and
of the dust emission from grains at a range of temperatures. From this, we can
calculate the distribution of galaxy properties at any redshift, including the
complete spectrum of each galaxy in the model from the far-UV to the sub-mm.

In this paper we compare the predicted properties for local galaxies with a
wide range of observational data. A future paper will be devoted to high-z
galaxies \citep{paper2}. In Sections~\ref{sec:galform} and \ref{sec:grasil} we
describe, respectively, the galaxy formation model and the spectrophotometric
model.  Section~\ref{sec:procedure} describes how we generate model galaxy
catalogues for both normal and starburst galaxies. The comparison with
observations (SEDs, extinction properties, colors, etc.) is presented in
Section~\ref{sec:compar} for spiral galaxies, and in Section~\ref{sec:compsb}
for starbursts. The model luminosity functions at different wavelengths are
compared with observations in Section~\ref{sec:lfs}. Section~\ref{sec:sfrcal}
uses the models to predict the relationship between the star formation rate and
the luminosities in various UV and IR bands, and to assess the accuracies of
these as star formation indicators. Section~\ref{sec:conc} presents a summary
and conclusions.

\section{Semi-analytical galaxy formation model}
\label{sec:galform}

\begin{table}[!htb] 
\begin{center}
\caption{Adopted values for GALFORM parameters. 
\label{table:galform}}
\begin{tabular}{lc}
\tableline
\multicolumn{2}{c}{{\em Cosmology}}\\
\tableline
$\Omega_0$ & 0.3 \\
$\Lambda _0$ & 0.7 \\
$h$ & 0.7 \\
$\Omega _{\mathrm{b}}$ & 0.02 \\
$\Gamma$ & 0.19 \\
$\sigma _8$ & 0.93 \\
\tableline
\multicolumn{2}{c}{{\em Star formation and feedback}}\\
\tableline
$\epsdisk$ & $6.7\times 10^{-3}$ \\
$\alphastar$ & -1.5 \\
$\alphahot$ & 2.0 \\
$\Vhot$ (km/s) & 150.0 \\
\tableline
\multicolumn{2}{c}{{\em Stellar populations}}\\
\tableline
IMF & \citet{kenn83} \\
$\Upsilon$ & 1.4 \\
$p$ & 0.02 \\
$R$ & 0.29 \\
\tableline
\multicolumn{2}{c}{{\em Mergers and bursts}}\\
\tableline
$\fellip$ & 0.3 \\
$\epsburst$  & 0.5\\
$\eta$  & 0.1\\
\tableline
\end{tabular} 
\end{center}
\end{table}

We calculate the formation histories and global properties of galaxies using
the semi-analytical galaxy formation model (GALFORM) of \citet{cole99}, a
development of that described in \citet{cole94} and \citet{baugh98}. The
principle of the model is to calculate the formation and evolution of dark
matter halos starting from an assumed cosmology and initial spectrum of density
fluctuations, and then to calculate the evolution of the baryons (gas and
stars) within these evolving halos using a set of simple, physically-motivated
rules to model gas cooling, star formation, supernova feedback and galaxy
mergers. We describe here only the main features of the model, and refer the
reader to \citet{cole99} for more details and for a discussion of the effects
of varying parameters with respect to standard values given in
Table~\ref{table:galform}.

\paragraph{(a) Cosmology:} The cosmology is specified by the present-day
density parameter $\Omega_0$, cosmological constant $\Lambda_0$, baryon
fraction $\Omega_b$ (all in units of the critical density) and the Hubble
constant $H_0=100h\ \kms \Mpc^{-1}$. We assume a cold dark matter (CDM) model,
with the initial spectrum of density fluctuations having shape parameter
$\Gamma$ and amplitude $\sigma_8$ (the r.m.s.\ density fluctuation in a sphere
of radius $8 h^{-1} \Mpc$).

\paragraph{(b) Halo evolution:} Dark matter halos form through a process of
hierarchical clustering, building up through merging from smaller objects. At
any cosmic epoch, we calculate the number density of halos as a function of
mass from the Press-Schechter (\citeyear{ps}) formula. We then calculate halo
merger histories, describing how a halo has formed, for a set of halos of
different masses, using a Monte-Carlo algorithm based on the extended
Press-Schechter formalism. We generate many different realizations of the
merger history for each halo mass. We then follow the process of galaxy
formation separately for each of these realizations.

\paragraph{(c) Cooling and collapse of gas in halos:} Diffuse gas is assumed to
be shock-heated to the virial temperature of the halo when it collapses, and to
then cool radiatively out to a radius determined by the density profile of the
gas and the halo lifetime. The gas which cools collapses to form a rotationally
supported disk, for which the half-mass radius $\rdisk$ is calculated
assuming that dark matter and associated gas are spun up by tidal torques, and
that the angular momentum of the gas is conserved during the collapse. The gas
supply by cooling is assumed to be continuous over the lifetime of the halo.

\paragraph{(d) Star formation in disks:} Stars form from the cold gas in
the disk, at a rate 
\begin{equation}
\psi=\Mcold/\taustdisk,
\end{equation}
where the star formation timescale is assumed to be 
\begin{equation}
\taustdisk = \epsdisk^{-1} \taudisk \left(\Vdisk/200 \kms \right)^{\alphastar}
\end{equation}
where $\Vdisk$ is the circular velocity at the
half-mass radius of the disk, and $\taudisk=\rdisk/\Vdisk$ is
the dynamical time. $\epsdisk$ is the fraction of gas
converted into stars in one dynamical time, for a galaxy with circular
velocity $\Vdisk=200 \kms$.
The scaling of the star formation timescale with dynamical time is
motivated by observations of star formation in nearby galaxies
\citep{kenn98}, but modified to reproduce the observed dependence of
gas fraction on luminosity.

\paragraph{(e) Supernova feedback in disks:} The energy input from
supernovae is assumed to reheat gas in the disk and eject it into the
halo at a rate 
\begin{equation}
{\dot M}_{\rm eject} = \betadisk\psi, 
\end{equation}
where for $\betadisk$ we assume 
\begin{equation}
\betadisk = \left( \Vdisk/\Vhot\right)^{-\alphahot}
\label{eq:disk-feedback}
\end{equation}
Gas which has been ejected is assumed to be unavailable for cooling
until the halo has doubled in mass through merging.
The motivation for the this parameterization is that the rate of gas
ejection should be proportional to the rate of supernovae, and also
depend on the escape velocity from the disk, which in turn is related
to the circular velocity. Our standard case $\alphahot=2$ is
equivalent to the assumption that a constant fraction of the Type~II
supernova energy goes into ejecting gas from the disk, if the escape
velocity is proportional to $\Vdisk$.

\paragraph{(f) Galaxy mergers and morphology:} The galaxy
morphology (\ie whether it is a spiral or elliptical) is determined by
merging. Following the merger of two halos, the largest pre-existing
galaxy is assumed to become the central galaxy in the new halo, while
the other galaxies become satellite galaxies. The central galaxy can
continue to grow a disk by cooling of gas from the halo. The satellite
galaxies merge with the central galaxy on a timescale equal to that
for dynamical friction to make the orbits decay. The merger is classed
as a {\em major merger} if the mass ratio of the satellite to central
galaxy exceeds a value $\fellip$, and as a {\em minor merger}
otherwise. In a {\em major merger}, any pre-existing stellar disks are
destroyed, producing a stellar spheroid (elliptical galaxy or bulge),
and any remaining cold gas is consumed in a burst of star
formation. The star formation timescale in the burst is related to the
dynamical time of the bulge as described below. The spheroid can grow
a new disk by cooling of halo gas. In a {\em minor
merger}, the stars from the satellite galaxy add to the bulge of the
central galaxy, while the cold gas adds to the disk, but no burst is
triggered. In either case, the half-mass radius $\rbulge$ of the
spheroid produced in a merger is calculated using an energy conservation
argument. Galaxies are classified into different morphological types
based on their bulge-to-disk luminosity ratios.

\paragraph{(g) Star formation and feedback during bursts:} As
already mentioned, star formation bursts are assumed to be triggered
by major mergers of galaxies. In \citet{cole99}, these bursts were
modelled in a very simple way, with the conversion of gas into stars
being assumed to be instantaneous, since the galaxy properties examined
there were not sensitive to the detailed time dependence. In this
paper, we model the bursts in more detail. We assume that star
formation during bursts follows a law analogous to that for star
formation in disks:
\begin{equation}
\psi=\Mcold/\taustburst,
\end{equation}
with star formation timescale  
\begin{equation}
\taustburst = \epsburst^{-1} \taubulge 
\label{eq:tauburst}
\end{equation}
where $\taubulge = \rbulge/\Vbulge$ is the dynamical time of
the spheroid formed in the merger, $\Vbulge$ being the circular
velocity at $\rbulge$. As in \citet{cole99}, feedback is modelled
as in disks except with $\Vbulge$ replacing $\Vdisk$ in
eqn.(\ref{eq:disk-feedback}), assuming the same values for $\Vhot$
and $\alphahot$, giving a feedback factor $\betaburst$. Since we
assume that no new gas is supplied by cooling during the burst, the
star formation rate and cold gas mass decay during the burst as
$\exp(-t/\tau_e)$, where 
\begin{equation}
\tau_e = \taustburst/(1-R+\betaburst),
\label{eq:taue}
\end{equation}
and $R$ is the recycled fraction, discussed below. The burst is assumed
to occur in a region of half-mass radius $\rburst$, where
\begin{equation}
\rburst = \eta\, \rbulge
\end{equation}

More details on the geometry assumed for starbursts are given in
\S~\ref{sec:grasil}. For simplicity, the metallicity of the gas in the burst
and of the stars formed during the burst are taken to be constant, and equal to
the mean metallicity of the stars formed during the burst as calculated by the
GALFORM model. The star formation in a burst is truncated at a time $5\tau_e$
after the burst began, \ie after 99\% of the gas in the burst has either been
converted into stars or blown out of the galaxy by supernova feedback. At this
time, the remaining gas and dust in the burst region are assumed to be
dispersed. Star formation then starts again in a normal galactic disk
surrounding the bulge, if one has formed by cooling of halo gas since the major
merger that triggered the burst.

\paragraph{(h) Chemical evolution:} 
We assume that a fraction $1/\Upsilon$ of the mass formed into
stars goes into visible stars ($0.1<m<125 M_{\odot}$), while the
remainder goes into brown dwarfs ($m<0.1 M_{\odot}$). For visible
stars we adopt a universal 
IMF, similar to that in the solar
neighbourhood. In \citet{cole99} and in this paper, we
use the form proposed by \citet{kenn83}, which is consistent with the
``best estimate'' of \citet{scalo98}:
\begin{eqnarray}
dN/d\ln m & \propto & m^{-0.4} \qquad (m<1 \Msol) \nonumber \\
          & \propto & m^{-1.5} \qquad (m>1 \Msol) 
\label{eq:imf}
\end{eqnarray}
We use the instantaneous recycling approximation to calculate the
evolution of the abundance of heavy elements of the cold gas
($\Zcold$) and stars ($\Zstar$) in each galaxy, together with that of
the hot gas in the halo ($\Zhot$), including the effects of inflows
and outflows between the galaxy and halo. The chemical evolution
depends on the recycled fraction $R$ and the yield of heavy elements
$p$.

\paragraph{(i) Stellar population synthesis and dust extinction:}
In \citet{cole99}, we calculated the luminosity evolution of each
galaxy at different wavelengths using the stellar population synthesis
models of \citet{bc99}. The effects of dust {\em extinction} were
calculated in a simple way using the dust models of \citet{ferrara99},
which assume a smooth (unclumped) distribution for both the dust (in a
disk) and stars (in a disk and a bulge). In the present paper, we use
instead the combined stellar population and dust model GRASIL
\citep{silva98} to calculate the galaxy luminosities and spectra
including both {\em extinction} and {\em emission} by dust. The
stellar population part of GRASIL is similar to the Bruzual
\& Charlot model, as both are based on similar stellar evolution
tracks and stellar spectra. The dust part of GRASIL is however
considerably more sophisticated than the Ferrara \etal models, in that
GRASIL allows for clumping of both dust and stars, and calculates the
grain heating and emission as well as the extinction.

\paragraph{} The parameters we have chosen for the GALFORM model are the same
as those of the standard $\Lambda CDM$ model of \citet{cole99}, apart
from $\epsburst$ and $\eta$ describing the timescale and radius of
bursts, which were not considered in Cole \etal. These parameters
values are given in Table~\ref{table:galform}, and were obtained by
comparing the model to observations of nearby galaxies, without any
consideration of the far-IR or UV properties. We refer the reader to
Cole \etal for a complete discussion of the effects of varying the
'old' GALFORM paramaters, and for a systematic presentation of the
influence of these parameters on the optical-NIR properties of
galaxies (LFs, Tully-Fisher relation, disk sizes, morphology, gas
content, metallicity, M/L ratios and colours). We only recall here the
main observational constraints used to fix each of these parameters:
$\epsdisk$ - gas fraction of $L_\star$ galaxies; $\alphastar$ -
variation of gas fraction with luminosity; $\alphahot$ - faint end of
LF and Tully-Fisher relation ; $\Vhot$ - faint end of LF and sizes of
low-$L$ spirals; IMF - observations of solar neighbourhood; $\Upsilon$
- $L_\star$ in LF; $p$ - metallicity of $L_\star$ ellipticals;
$\fellip$ - morphological mix of $L_\star$ galaxies.

Values for $\epsburst$ and $\eta$ are instead obtained later in this paper by
detailed comparison of the results of the combined GALFORM+GRASIL models with
observed properties of bursting galaxies.


\section{The Stellar Population and Dust Model}
\label{sec:grasil}

Far-UV to mm SEDs of model galaxies are calculated using the GRASIL
code \citep{silva98}, which follows both the evolution of stellar populations
and absorption and emission by dust. GRASIL calculates the following: (i)
emission from stellar populations; (ii) radiative transfer of starlight through
the dust distribution; (iii) heating and thermal equilibrium of dust grains (or
thermal fluctuations for small ones); and (iv) emission by dust grains.

\subsection{Stellar Population Model}

The single stellar population (SSP) libraries included in GRASIL are
based on the Padova stellar models and cover a large range in age and
metallicity. They include the effects of dusty envelopes around AGB
stars \citep{bress98}. The age and metallicity distribution of a
composite stellar population is specified by the birthrate function
$\Psi(t,Z)$, where $\Psi(t,Z)\,dt\, dZ$ gives the mass of stars that
were formed in the time interval $(t,t+dt)$ with metallicities in the
range $(Z,Z+dZ)$. The SED for the composite stellar population at time
$t$ is then obtained using
\begin{equation}
L_{\lambda}(t) = \int_0^t dt' \int_0^1 dZ\; l_{\lambda}(t-t',Z)
\Psi(t',Z)
\label{eq:sspsum}
\end{equation}
where $l_{\lambda}(\tau,Z)$ is the SED of a SSP of age $\tau$ and
metallicity $Z$ for the assumed IMF.

For our semi-analytical galaxy formation model, $\Psi(t,Z)$ is
calculated for each galaxy by summing over all the progenitor galaxies
which have merged to produce that galaxy, separately for the disk and
bulge components. The progenitor galaxies each had their own star
formation and chemical history, so that the composite $\Psi(t,Z)$
obtained in general has a broad distribution of metallicity at each
each age, \ie there is no unique age-metallicity relation $Z(t)$.

\begin{figure}[!htb] 
\epsscale{1.0}
\plotone{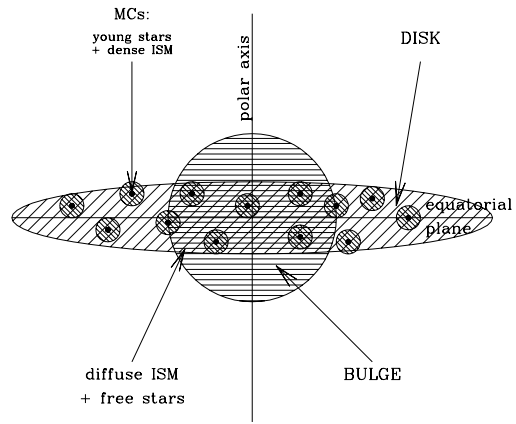} 
\caption{Sketch of geometry of stars and dust in the GRASIL model.}
\label{fig:cartoon2}
\end{figure}

\begin{table}[!htb] 
\begin{center}
\caption{Adopted values for the adjustable GRASIL parameters. 
Note that the results depend on the ratio $M_c/r_c^2$ rather
than on the two quantities individually (see \S~\ref{par:geodus}).
\label{table:dust}}
\begin{tabular}{lc}
\tableline
$h_{z}/h_{R}(disk)$ & 0.1 \\
$h_{z}/h_{R}(burst)$ & 0.5 \\
$h_{z}(dust)/h_{z}(stars)$ & 1 \\
$f_{mc}$ & 0.25 \\
$M_c/r_c^2$ &  $10^6\Msol/(16\pc)^2$ \\
$\tesc(disk)$ & 2\Myr \\
$\tesc(burst)$ & 10\Myr \\
\tableline
\end{tabular} 
\end{center}
\end{table}

\subsection{Dust Model}
\label{sec:grasil_dust}

GRASIL computes the radiative transfer of starlight, the heating of dust
grains, and the emission from these grains with a self-consistent calculation
of the distribution of grain temperatures, for an assumed geometrical
distribution of the stars and dust, and a specific grain model. The dust is
divided into two components, dense molecular clouds and diffuse cirrus in a
disk. Stars form inside clouds and progressively leak out.

The details are given in \citet{silva98}, but for convenience we summarize the
main features here, focusing on the modifications introduced for the purposes
of this application. Those GRASIL parameters which are not provided by GALFORM,
and are in this sense additional adjustable parameters of the combined
GALFORM+GRASIL semi-analytic modelling, are listed in Table~\ref{table:dust},
together with the adopted values for our standard case. See
Fig.~\ref{fig:cartoon2} for a sketch of the geometry of our model.

\paragraph{(a) Geometry of stars:} 

The stars are in two components (Silva et al.\ (1998) considered only pure disk
and pure bulge systems): (i) a spherical bulge with an analytic King model
profile, $\rho \propto (r^2+r_c^2)^{-3/2}$ for $r<r_t$, with concentration
parameter $\log(r_t/r_c)=2.2$; (ii) a disk with a radially and vertically
exponential profile, scalelength $h_R$ and scaleheight $h_z$. As described in
\S~\ref{sec:galform}, the disk and bulge masses, $\Mdisk$ and $\Mbulge$, and
half-mass radii, $\rdisk$ and $\rbulge$, for any galaxy are predicted by the
galaxy formation model. The bulge core radius is related to the bulge half-mass
radius by $r_c = \rbulge/14.6$, while the disk scalelength $h_R$ is related to
the disk half-mass radius by $h_R = \rdisk/1.68$. The star formation histories
are also calculated separately for the disk and bulge by GALFORM. However, the
disk axial ratio $h_z/h_R$ is a free parameter of the GRASIL model.

As partially anticipated in \S~\ref{sec:galform}, in galaxies undergoing
bursts, the burst star formation, as well as the gas and dust, are assumed to
be in an exponential disk, but with half-mass radius $\rburst = \eta\ \rbulge$
rather than $\rdisk$. The axial ratio $h_z/h_R$ of the burst region is allowed
to be different from that for disks in non-bursting galaxies. The stars which
were formerly in the disks of the galaxies before the galaxy merger which
triggered the burst are assumed to become part of the bulge following the
merger.

\paragraph{(b) Geometry of gas and dust:} \label{par:geodus} The gas and dust
are in an exponential disk, with the same radial scalelength as the disk stars
(either $\rdisk$ for normal galaxies or $\rburst = \eta\ \rbulge$ for
starbursts), but in general with a different scaleheight, so that
$h_{z}(dust)/h_{z}(stars)$ is a free parameter. The gas and dust are in two
components within the disk, molecular clouds and the diffuse ISM. The latter
corresponds to the cirrus dust.  The total gas mass $\Mcold$ and its
metallicity $\Zcold$ are calculated by the galaxy formation model, but the
fraction of the gas in clouds, $f_{mc}$, and the cloud mass $M_c$ and radius
$r_c$ are free parameters of GRASIL, though the results actually depend only on
their combination $M_c/r_c^2$, which determines, together with the dust/gas
ratio (see point (d) below), the optical depth of the clouds \citep{silva98}.

\paragraph{(c) Young stars and molecular clouds:}
Stars are assumed to form inside the molecular clouds, and then to
escape on a timescale $\tesc$. Specifically, the fraction of stars
still inside clouds at time $t$ after they formed is assumed to be
given by
\begin{eqnarray}
F(t) & = & 1 \qquad(t<\tesc) \nonumber \\
     & = & 2-t/\tesc \qquad(\tesc<t<2\tesc) \nonumber \\
     & = & 0 \qquad(t>2\tesc) 
\label{eq:tesc}
\end{eqnarray}
We allow $\tesc$ to take different values in normal disks and in
bursts, in keeping with the results of \citet{silva98}. Indeed, given
the small size scale and the intensity of the star formation activity
in bursts, it is conceivable that the star-forming environment is
quite different from that in normal spiral galaxies (see also
\S~\ref{sec:choice}).

\paragraph{(d) Dust grain model:}
The dust is assumed to consist of a mixture of graphite and silicate
grains and polycyclic aromatic hydrocarbon molecules (PAHs), each with
a distribution of grain sizes. Absorption and emission properties are
calculated for each grain composition and size. The grain mix and size
distribution were chosen by \citet{silva98} to match the extinction
and emissivity properties of the local ISM, and are not varied
here. The dust/gas ratio $\delta$ in the clouds and diffuse ISM is
assumed to be proportional to the gas metallicity, with a value
$\delta= 1/110$ for $Z=\Zsol=0.02$. Thus, the total dust mass in a
galaxy scales as $\Mdust \propto \Zcold \Mcold$.

\paragraph{(e) Radiative transfer, dust heating and re-emission:} The
luminosities of the different stellar components (bulge stars, disk
stars, and young stars still in clouds) are calculated using the
population synthesis model described above. The GRASIL code then
calculates the radiative transfer of the starlight through the dust
distribution. Whilst in molecular clouds a full radiative transfer
calculation is performed, the effects of scattering by diffuse dust
are included only approximately, by assuming that the effective
optical depth for absorption is related to the true absorption and
scattering optical depths $\tau_{abs}$ and $\tau_{scat}$ by
$\tau_{abs,eff} = \sqrt{\tau_{abs} (\tau_{abs} + \tau_{scat})}$. Thus
the dust-attenuated stellar radiation field can be calculated at any
point inside or outside the galaxy.  Then GRASIL calculates for each
point in the galaxy the absorption of radiation, thermal balance and
re-emission for each grain composition and size.  Thus, the
distribution of grain temperatures is calculated self-consistently for
the assumed geometry of the stars and dust, including the effects of
temperature fluctuations for small grains. The final galaxy SED
$L_{\lambda}$ is obtained by adding the contributions from the
starlight (attenuated by dust) and from the dust re-emission, and
depends on the inclination angle at which the galaxy is
viewed. Emission from dust in the envelopes of AGB stars is included
in the SSPs.

Our computations allow us to calculate the amount of energy emitted in the PAH
bands, but theoretical predictions of the detailed shapes of the emission
features are rather uncertain. Therefore we use the Lorentzian analytical fits
to the observed PAH profiles for the Ophiuchus molecular cloud from
\citet{boulanger98}.

\subsection{Choice of GRASIL adjustable parameters and new GALFORM parameters}
\label{sec:choice}

The values of GRASIL parameters (Table~\ref{table:dust}) not provided by
GALFORM have been based on a variety of observational data for galaxies in the
local universe. For some of them, the choices were made by trying to match
model predictions to the observational data, as is discussed in more detail in
the relevant sections of this paper. We now summarize the reasons for these
choices and for those of the two GALFORM parameters ($\epsburst$ and $\eta$)
not considered in \citet{cole99}

\paragraph{(a) $\epsburst$:} this is chosen mainly so as to reproduce the
bright end of the IR luminosity function, which is dominated by bursts
triggered by galaxy mergers (\S\ref{sec:irlf}). A secondary (weak) constraint
is to reproduce the relation between $L_{IR}/L_{UV}$ and total luminosity or UV
slope $\beta$ observed for starburst nuclei (\S\ref{sec:burst}). The value
controls both the luminosity and lifetime (and thus number density) of
starbursts.

\paragraph{(b) $\eta = \rburst/\rbulge$:} the choice of this is mainly based on
the observational fact that starburst activity is usually confined to a nuclear
region with a size much smaller than the galaxy as a whole, by about one order
of magnitude \citep[\eg][and references therein]{sm96}. For instance, in Arp
220 most of the molecular gas is found in the central $\sim 300\pc$
\citep{SYB97}, and the mid--IR light is dominated by more or less the same
region \citep{KBA92}, while the half-light radius for the old stellar
population is $\sim 3\kpc$ \citep{wright90}. The value of $\eta$ controls the
amount of extinction of starlight from bursts by the diffuse ISM, which however
is usually overwhelmed (in bursts) by extinction in molecular clouds (see
\S\ref{sect:ext}). Therefore our results are not very sensitive to the precise
choice of this parameter, nor to the value of $h_z/h_R$ in starbursts
(discussed below).

\paragraph{(c) $h_z/h_R$:} for normal disks, we choose a value of 0.1,
consistent with observations of the stellar light distributions in edge-on
spiral galaxies \citep[\eg][]{xil99}. It is also the typical value used by
\citet{silva98} to fit the SEDs of spiral galaxies. This value is also
important, and the adopted value turns out to be suitable, to match the
observed difference in extinction between spiral galaxies seen edge-on and
face-on (\S\ref{sec:ext}). Apart from this test, most predicted properties are
not very sensitive to $h_z/h_R$. The choice of $h_z/h_R=0.5$ for starbursts is
based on general observational indications that they are only moderately
flattened.

\paragraph{(d) $h_z(dust)/h_z(stars)$:} this parameter has a significant effect
on how much starlight is absorbed in the diffuse medium. From observations of
our own galaxy it is known that the scaleheight of stars increases with the age
of the stellar population, so that there is no unique value for
$h_z(dust)/h_z(stars)$. The scaleheight of the gas is comparable to that of the
youngest stars. Since we are particularly interested in having a realistic
estimate of the extinction in the UV, both because it is strongest there and
because this is an important source for dust heating, we choose
$h_z(dust)/h_z(stars)=1$ to match what is seen for the young stars.

\paragraph{(e) $f_{mc}$:} this can be estimated observationally from the ratio
of molecular to atomic hydrogen in galaxies, since in normal spiral galaxies
most of the hydrogen in molecular clouds is in $H_2$, while most of the
intercloud medium is atomic $HI$. Our adopted $f_{mc}$ implies a ratio $H_2/HI$
similar to the typical one for $L_\star$ spirals found by \citet{sage93}.
Larger values reduce the extinction in the diffuse ISM and produce a somewhat
colder molecular clouds emission, but our results are in general not
significantly affected as long as we keep $f_{mc}$ in the range 0.2--0.8.

\paragraph{(f) $M_c$,$r_c$}: as already remarked (\S~\ref{par:geodus}) the
predicted SEDs depend on the ratio $M_c/r_c^2$, rather than $M_c$ and $r_c$
separately. Thus $M_c$ has been chosen to match typical giant molecular clouds
in our own and nearby galaxies, while $r_c$ is chosen based on the results of
\citet{silva98}, who tuned $M_c/r_c^2$ to fit the SEDs of starburst galaxies in
particular. The resulting value for $r_c$ is consistent with direct
measurements of cloud radii.

\paragraph{(g) $\tesc$:} this is a very important parameter in the model, since
it is this that mainly controls how much of the radiation from young
stellar populations is absorbed by dust. \citet{silva98} found from
detailed fits to 3 nearby spirals values of 2.5, 3 and 8 Myr. For
normal spirals, we favor a value of 2 Myr, close to the lower limit of
this range, rather than the average 5 Myr. Although the latter
provides an equally good overall fit to the LFs (somewhat better for
IRAS colors and LFs, \S~\ref{sec:irlf}, but somewhat worse for the UV
LF), the former is more consistent with the massive star census in our
own and nearby galaxies, which suggests that the time for which the
stars are obscured by dust is about the 20 \% of the total lifetime
for the brightest stars, above say 30 $\Msol$, whose lifetime is
around 6 Myr. For starbursts, the value we choose is based mainly on
the comparison with properties of UV-bright starbursts in
\S\ref{sec:burst}. This leads us to a value closer to that of normal
spirals than the values $\tesc = 20$-$60\ \Myr$ found by
\citet{silva98} from fitting 3 nearby starbursts, and suggests that
the starburst galaxies used by Silva \etal may be not representative
of the whole population. The difference could also be due in part to
the more complex geometry adopted in this paper for starburst
galaxies.

\section{Generation of model galaxy catalogues} 
\label{sec:procedure}

The GALFORM model is run for a set of dark matter halos covering a
large range in mass, and generates a catalogue of model galaxies,
including information about the following properties for each galaxy
at the chosen epoch: stellar masses $\Mdisk$, $\Mbulge$, and half
mass radii $\rdisk$ and $\rbulge$ of the disk and bulge, mass
$\Mcold$ and metallicity $\Zcold$ of gas in the disk, and the star
formation histories $\Psi(t,Z)$ of the disk and bulge separately,
including both star formation in disks and during bursts, and
specifying the metallicity distribution of the stars of each age. In
addition, each galaxy has a weight or number density $n$, such that
that galaxy should appear $N=nV$ times in an average volume of the
universe $V$.

\clearpage 

The GALFORM code outputs all the galaxies for each different halo that is
calculated, down to a minimum mass controlled by the mass resolution of the
merger tree. In practice, this means that the model catalogue contains many
more low mass galaxies than high mass galaxies. Running the GRASIL code on
every galaxy in the original catalogue is neither feasible (because of computer
time) nor necessary. We therefore select a subset of galaxies from the
catalogue chosen to sample galaxies more evenly in mass, and redistribute the
weights to give the same total number density in each mass range. The GRASIL
code is then run on each galaxy in this reduced catalogue to give the SED
$L_{\lambda}$ including both stellar emission and dust absorption and emission,
and statistical properties (\eg luminosity functions) are then calculated
making use of these weights. In fact, we calculate 2 samples of galaxies, a
``normal'' sample and a ``burst'' sample, as follows:


\noindent{\em(a) Normal galaxies:} By ``normal'' galaxies, we here simply mean
galaxies not selected to have had a recent burst. From the parent GALFORM
catalogue, we select a sample with equal numbers of galaxies in equal bins in
$\log M_*$, $M_*$ being the total stellar mass of the galaxy. Within each mass
bin, galaxies are randomly selected (allowing for multiple selection of the
same galaxy) with probability proportional to their weight $n$. The selected
galaxies are then assigned new weights $n_i$, such that each galaxy within the
same bin has the same weight (multiply selected galaxies being counted as
separate objects), and that the sum of the weights (\ie number densities)
within a bin is the same as in the parent catalogue. We have used bins with
$\Delta\log M_*=0.3$ and about 40 galaxies per bin.

\noindent{\em(b) Burst galaxies:} By ``burst'' galaxies we mean
galaxies which have had a burst in the recent past, at whatever
redshift we are looking.  Bursts have short durations compared to the
age of the universe, so the fraction of galaxies undergoing a burst at
any one time is very small, but they can be very luminous, and so may
dominate the galaxy luminosity function at the highest
luminosities. In practice, our ``normal galaxy'' catalogue contains
too few galaxies in total to provide a representative sample of
galaxies seen during their burst phase. Rather than use a greatly
enlarged ``normal galaxy'' sample, it is more efficient to calculate a
separate sample of ``burst'' galaxies, as follows: for a redshift $z$,
we choose a subsample of galaxies which have had bursts during the
time interval $t(z)>t>t(z)-T$, where $t(z)$ is the age of the universe
at redshift $z$, with equal numbers of galaxies in equal bins in $\log
\Mburst$, $\Mburst$ being the mass of stars formed in the most recent
burst. The galaxies are assigned new weights $n_i$ analogously to the
case of normal galaxies, but now conserving the total number density
in bins of $\Mburst$ for the galaxies which have had bursts more
recently than $T$. For each burst galaxy, we then run GRASIL to
calculate the total galaxy luminosity at a set of times after the
start of the burst, chosen to sample all phases of the burst
evolution, including the highest luminosity phase of short
duration. If $T<<t(z)$, then the rate of bursts per unit volume during
the time interval $T$ can be taken as constant. Then, for the $i$th
galaxy in the $j$th phase in the burst evolution that lasts a time
$\Delta t_j$, the number density of galaxies that should be found in
this phase is \begin{equation} n_{ij} = n_i \left( \frac{\Delta
t_j}{T} \right) \end{equation} These weights can then be used to
calculate statistical properties such as luminosity functions. When
combining the ``normal'' and ``burst'' galaxy samples, the normal
galaxies with bursts more recent than $T$ are explicitly excluded, to
avoid statistical double-counting. In practice, we chose $T=t(z)/20$
at all $z$, with bins $\Delta\log \Mburst=0.3$, around 10 galaxies per
bin, and around 10 output times per galaxy, for $0<t-t_{burst} \lsim
100\tau_e$. For many calculations of statistical distributions, we
then interpolate between these output times to have more burst phases.

\begin{figure*}[]
\epsscale{2.0} 
\plotone{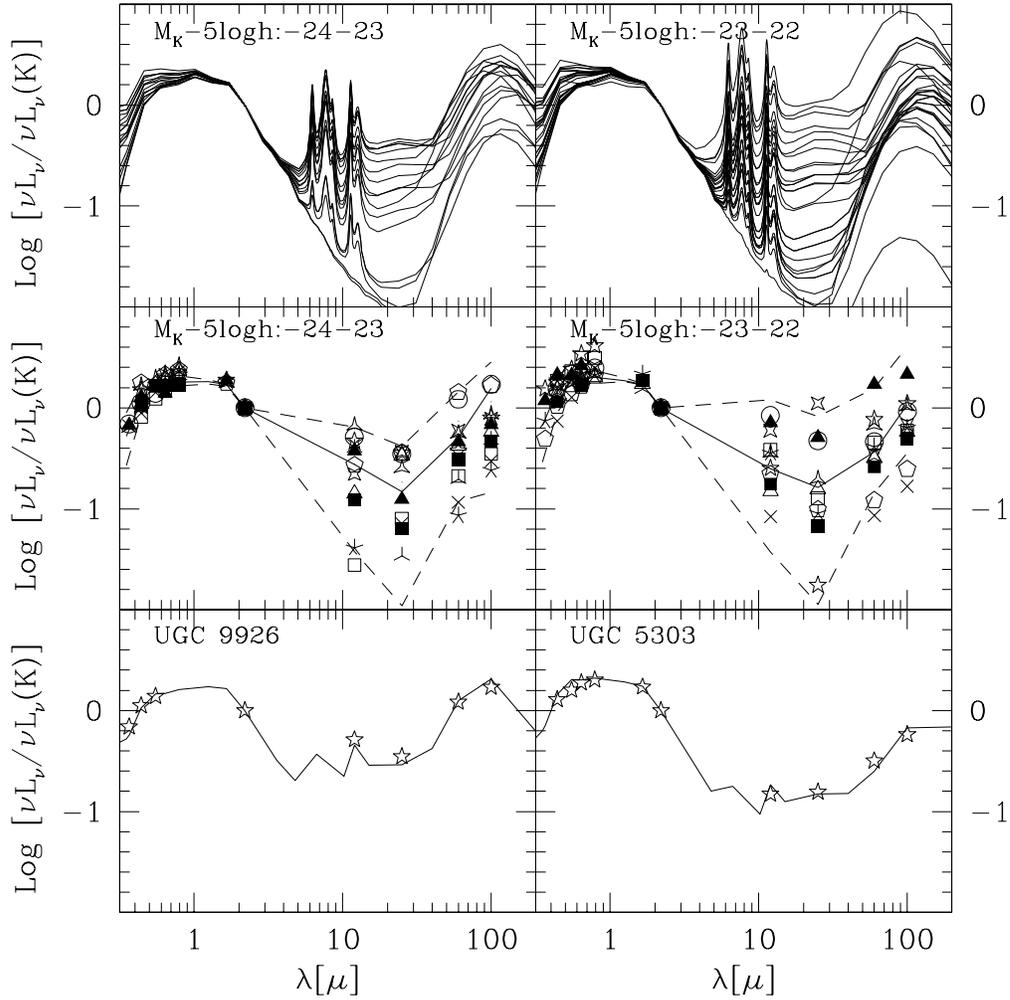}
\caption{Upper panels: SEDs (normalized to the K-band) of a
representative sample of spiral galaxies from the models. Middle
panels: Comparison of the models with the observed SEDs of a
sample of face-on spiral galaxies obtained by \citet{djvdk94}. The
median and the 10\% and 90\% percentiles of the model SEDs are plotted
as lines.  Lower panel: comparisons between observed SEDs of
individual galaxies from the sample and selected model SEDs. The
left-hand panels are for galaxies in the luminosity range
$-23>M_K-5\log h >-24$, and the right-hand panels for $-22>M_K-5\log h
>-23$.}
\label{fig:fkm4}
\end{figure*}

\begin{figure}[]
\epsscale{0.9} 
\plotone{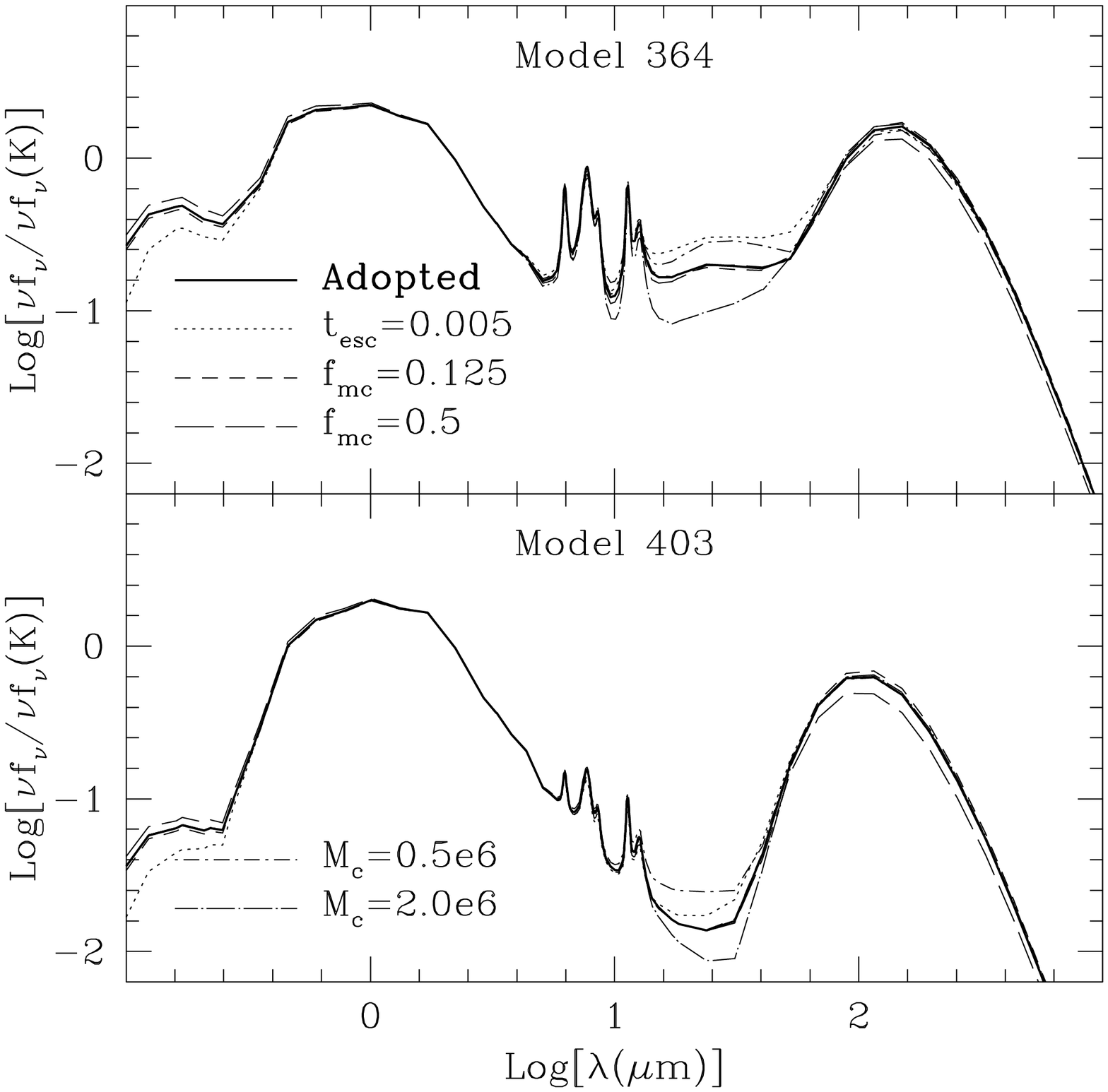}
\caption{Effects on typical SEDs of factor 2 variations of $f_{mc}$,
$M_c$ (effectively, the cloud optical depth) and $\tesc$. In each case
only one parameter is varied with respect to the standard model.}
\label{fig:compmod}
\end{figure}

\section{Properties of spiral galaxies}
\label{sec:compar}

In this section, we test the model predictions for disk galaxies against
observed emission and absorption properties of nearby spirals.

\subsection{SEDs of face-on spirals}

We compared the predicted near-UV to far-IR SEDs of our model galaxies with the
broad-band SEDs of a complete sample of nearby spiral galaxies \citep{djvdk94},
consisting of a diameter-limited sample of 86 nearly face-on, disk-dominated
galaxies. \citeauthor{djvdk94} measured fluxes of these galaxies in the BVRIHK
bands. We have supplemented these with U-band magnitudes from the literature
and IRAS $12,25,60,100 \mum$ fluxes from \citet{saunders97}.

We considered only those model galaxies with bulge to total light
ratio $B/T\le 0.5$ in the B-band, corresponding to the range of types
in the de~Jong sample \citep[\eg][]{sv86}. From Figure \ref{fig:fkm4}
it is apparent that the models reproduce the observed spectral trends
reasonably well. This is particularly impressive since the ratio
between the infrared and the optical-UV spans more than one order of
magnitude, both in the observed and in the theoretical SEDs.  The
predicted infrared emission peaks at wavelengths somewhat larger than
those sampled by IRAS, in agreement with recent ISO observations
\citep[\eg][]{alton98}. The emission in the mid-infrared is dominated
by PAH molecular bands.

Figure \ref{fig:compmod} shows the effects on typical SEDs of factor 2
variations in the molecular cloud fraction $f_{mc}$, their mass $M_c$
(\ie their optical depth, having fixed the radius) and the escape
timescale $\tesc$. The effects are mostly confined to the mid-IR
between 8 and 40 $\mu$m and in the UV below 0.4 $\mu$m. In these
spectral regions, the predicted flux may change by up to a factor
$\sim 2$, while the effects are almost negligible elsewhere. In the
mid-IR the most important parameter is the cloud optical depth, while
in the UV the effects of $\tesc$ dominate.

\begin{figure}[!htb] 
\epsscale{1.0}
\plotone{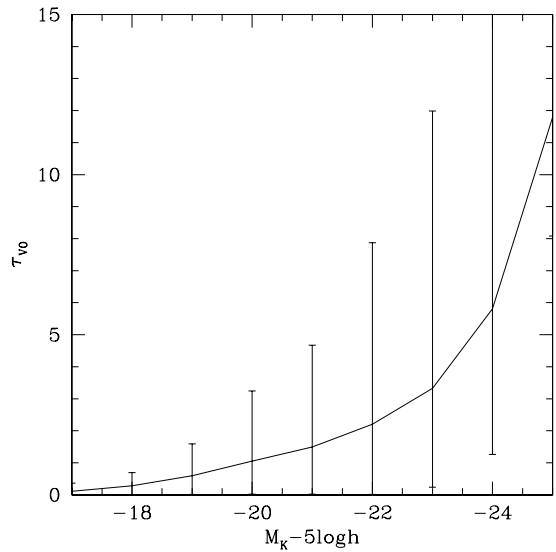}
\caption{Predicted dependence of dust optical depth on luminosity for
spiral galaxies in the model. $\tau_{V0}$ is the central face-on
extinction optical depth in the V-band for the diffuse dust component
in the disk. The line shows the median, and the error bars show the
10\% and 90\% percentiles. Model galaxies are selected to have
$B/T<0.4$ in the B-band.  }
\label{fig:tau_med}
\end{figure}

\clearpage 

\subsection{The global extinction in spiral galaxies}
\label{sec:ext}

The models predict that the extinction in galaxy disks should increase strongly
with galaxy luminosity, as shown in Figure~\ref{fig:tau_med}. Clearly, in
comparing predictions of dust extinction with observations, one must be careful
to specify the luminosity of the objects concerned.

There have been many attempts to measure or observationally constrain the total
dust extinction in galaxy disks, using a variety of techniques: the inclination
dependence of magnitudes or colours \citep[\eg][]{rc3,giovanelli95}, surface
brightness distributions in edge-on galaxies \citep[\eg][]{kylafis87}, colour
gradients in face-on disks \citep[\eg][]{dj96c}, and the ratio of far-IR to UV
luminosities \citep[\eg][]{xu95,buat96,wang96}. In general, different
techniques have given somewhat different answers.

\citet{xil99} estimate dust extinctions by fitting detailed models of the star
and dust distributions to the observed surface brightness distributions of
edge-on spiral galaxies. Their dust models include scattering. From their six
Sb-Sc spirals with luminosities in the range $-17.5> M_B-5\log h>-19.0$, we
obtain a median central face-on extinction optical depth $\tau_{V0} =0.6$. In
comparison, for edge-on galaxies in our model in the same luminosity range
(after extinction), and with $0.1 <B/T <0.3$ in the B-band, we find a median
value $\tau_{V0} =2.2$, which is significantly larger. There could be several
reasons for this difference between the models and the observations: there may
be a problem with the Xilouris \etal method for deriving $\tau_{V0}$ from the
observations, or the Xilouris \etal sample may not be representative, or the
problem might be with our assumption that the dust and stars have the same
exponential scalelength. The extinction-inclination observational test
discussed next implies extinctions for edge-on galaxies in this luminosity
range which are at least as large as those predicted by our model.

We considered the dependence of the net extinction on the
inclination angle at which a galaxy is viewed. This has been studied in many
papers using different methods, most recently by \citet{tully98}, who also
summarize the results from the earlier studies. Tully \etal measure the
dependence of $B-K$, $R-K$ and $I-K$ colours on galaxy inclination at a given
K-band luminosity, the K-band being chosen to minimize extinction effects. They
have a complete sample of spirals covering a large range in luminosity, $-18.5
\gsim M_K-5\log h \gsim -24.5$. They find a strong luminosity dependence, with
a difference in B-band extinction between edge-on and face-on galaxies of about
2 mag for the brightest galaxies, and negligible for the faintest ones.

\citet{tully98} follow the usual practice and parameterize the
extinction relative to that for the galaxy seen face-on as
\begin{equation}
A_\lambda^{i-0} \equiv m_{\lambda}(i)-m_{\lambda}(0) = \gamma_\lambda \log(a/b)
\label{eq:inc1}
\end{equation}
where $\gamma_\lambda$ is a function of the passband.  The axial ratio
$a/b$ is assumed to be related to the inclination angle $i$ by 
\begin{equation}
\cos \ i = \sqrt{\frac{(b/a)^2-q^2}{1-q^2}}
\label{eq:inc2}
\end{equation}
where $i=0$ for a face-on system, and $q$ is the axial ratio of a
galaxy seen edge-on.

\begin{figure}[!htb] 
\epsscale{1.0}
\plotone{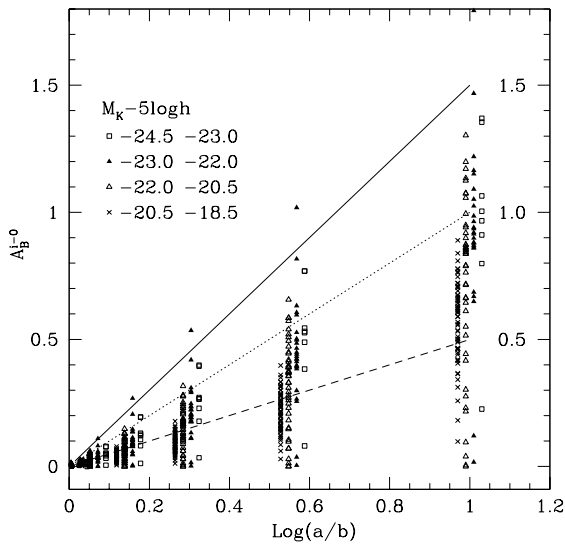}
\caption{Dependence of extinction on inclination angle predicted by our
galaxy models.  The quantity $A_B^{i-0}=M_B(i)-M_B(0)$ is plotted
against the apparent axial ratio $a/b$, for the Sa-Scd models (with
$0.04<B/T<0.4$ in the B-band). The galaxies are plotted with
different symbols in four K-band luminosity ranges, as detailed in the
figure.  The same galaxies are plotted seen at different
inclinations. For clarity, small horizontal offsets have been applied
to the galaxies in different luminosity ranges. Three lines are
plotted corresponding to equation~(\ref{eq:inc1}) with the slopes
$\gamma_B=$ 0.5, 1 and 1.5. }
\label{fig:inc}
\end{figure}

The models are compared with observations in Fig.~\ref{fig:inc}. We use
equation~(\ref{eq:inc2}) to convert from the model inclination angle to the
axial ratio, assuming $q=0.1$, which is the ratio $h_z/h_R$ adopted in our
galaxy models.  We considered model galaxies corresponding to the morphological
types Sa-Scd, and four ranges in K-band luminosity, corresponding to the ranges
chosen by \citet{tully98}, indicated by the different symbols in the figure.

The three lines in the figure correspond to different values of the
slope $\gamma_B$. The model galaxies approximately follow the linear
dependence on $\log(a/b)$ (equation~(\ref{eq:inc1})), but with slopes
$\gamma_B$ that are somewhat shallower, at any given luminosity, than
those observationally inferred by \citet{tully98}. For instance, for
the luminosity range $-23.0<M_K-5\log h<-22.0$, our models follow an
average slope $\langle \gamma_B \rangle \approx 0.9$, while Tully
\etal find $\gamma_B = 1.1 \pm 0.5$, after allowing for the K-band
extinction. The slope predicted by our models depends on the value
chosen for the parameter $h_z/h_R$ (see also \S~\ref{sec:choice}). We
have checked that increasing $h_z/h_R$ from our adopted value of 0.1
decreases the slope, while reducing it does not increase the slope
significantly.

The agreement for the inclination test is anyway acceptable. Part of the
discrepancy could be due to our simplified treatment of scattering by the
diffuse dust (\S~\ref{sec:grasil_dust}).  In our models, the absolute
extinction is often dominated by the molecular clouds, but the difference
between the face--on and the edge--on extinction is entirely due to the cirrus.
Comparisons of our model with that of \citet{ferrara99}, where the treatment of
scattering is more accurate, show that for the brightest objects this effect
can account for about 0.1-0.2 mag of the differential extinction in the B band.


\section{Properties of starburst galaxies}
\label{sec:compsb}

Starburst galaxies are broadly defined as galaxies in which the current star
formation rate is much greater than its time-averaged value, and the star
formation timescale correspondingly much shorter than the age of the universe.
This definition includes objects with a wide range of properties, from bursting
dwarf irregular galaxies \citep[\eg][]{thuan81} to the ultra-luminous IR
galaxies (ULIGs) found by IRAS \citep[\eg][]{sm96}.  In practice, a large
variety of observational criteria have been applied to select samples of
starburst galaxies, ranging from optical morphologies and spectra
\citep[\eg][]{balzano83} to IR colours and luminosities
\citep[\eg][]{armus90,lehnert95}. In our galaxy formation model, bursts are
assumed to occur following major mergers of galaxies, producing elliptical
galaxies from disk galaxies. For the ultra-luminous IR galaxies, the link
between the starburst activity and galaxy mergers is clearly established
\citep[\eg][]{sm96}, while for low-luminosity starbursts, additional triggering
mechanisms probably operate, which are not included in our model. In this
section we will compare the properties of starbursts predicted by our model
with those of various observational samples.

\clearpage 

\begin{figure*}[!htb]
\epsscale{2.0} 
\plotone{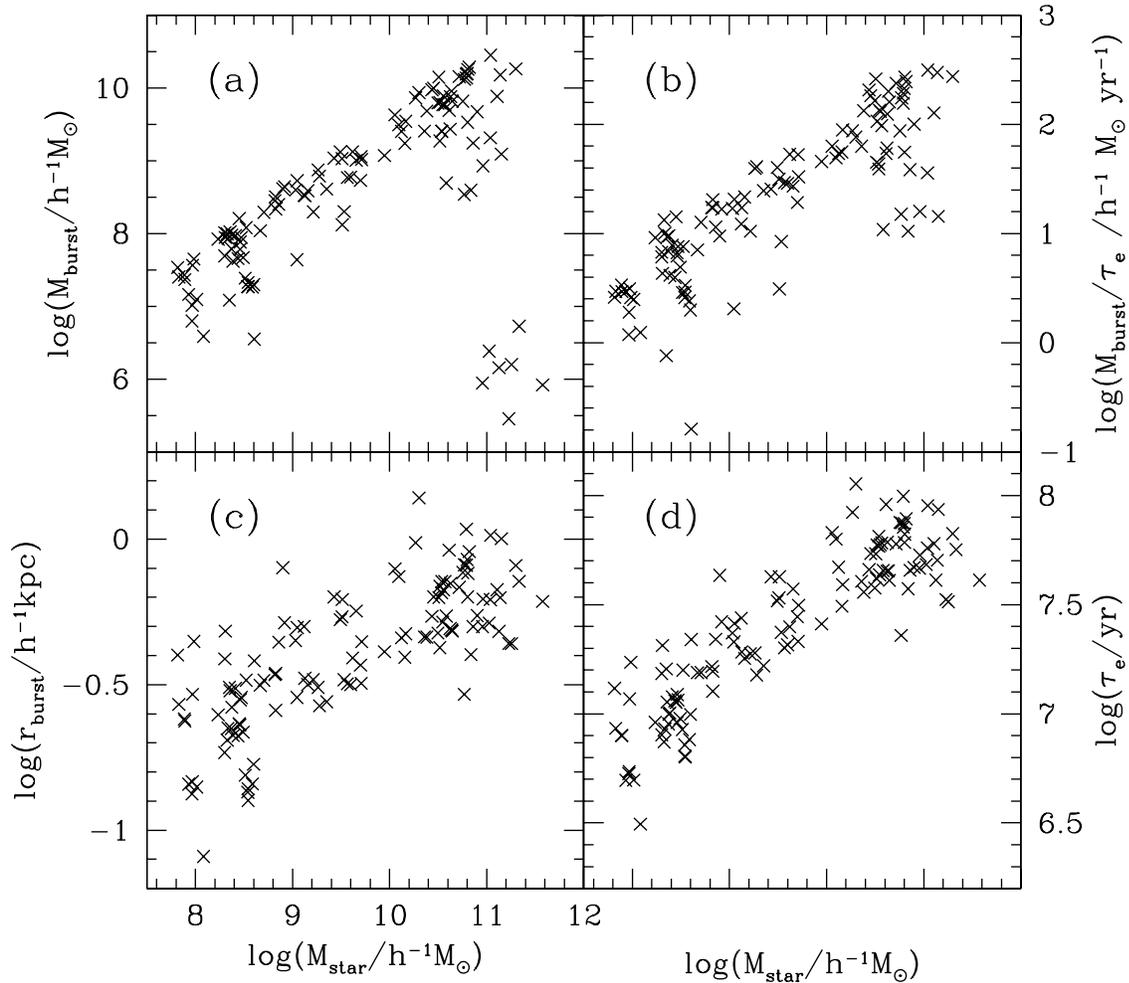}
\caption{Properties of starbursts in the model at $z=0$. (a) The mass of new
stars formed in the burst, $\Mburst$ vs the stellar mass of the
galaxy after the burst, $\Mstar$. (b) The peak star formation rate
in the burst $\Mburst/\tau_e$ vs $\Mstar$. (c) The half-mass radius
of the burst $\rburst$ vs $\Mstar$. (d) The exponential decay time
of the burst $\tau_e$ vs $\Mstar$.}
\label{fig:burstprops}
\end{figure*}

\subsection{Properties of starbursts in the model}

Figure~\ref{fig:burstprops} shows how various properties of the bursts
in our model vary with the stellar mass of the galaxy, $\Mstar$, after
completion of the burst. The total mass of new stars formed in the
burst, $\Mburst$, is seen to increase with the galaxy mass, with the
fraction of stars formed in the burst being typically between $\sim
1\%$ and $\sim 50\%$. An exception to this trend is the group of
points in the lower right corner of Fig.\ref{fig:burstprops}a
corresponding to small bursts occuring in large galaxies. These small
bursts are produced by mergers between gas-poor elliptical
galaxies. The main trend in panel~(a) is produced by mergers between
disk galaxies containing significant fractions of gas, and these
dominate the statistics at all burst masses. The star
formation rate during the burst is $(\Mburst/\tau_e) \exp(-t/\tau_e)$,
with $t$ measured from the start of the burst. The peak star formation
rate is thus $\Mburst/\tau_e$, and occurs at the beginning of the
burst. This peak SFR is seen also to increase with the host galaxy
mass. The half-mass radius $\rburst$ and exponential decay time
$\tau_e$ of the burst are assumed to scale with the half-mass radius
and dynamical timescale of the host galaxy, and also increase with
galaxy mass. Large bursts, with $\Mburst \sim 10^{10} h^{-1} \Msol$,
are predicted to occur in galaxies with $\Mstar\sim 10^{11}\Msol$, and
to have radii $\rburst\sim 0.5 h^{-1}\kpc$, star formation timescales
$\tau_e \sim 5\times 10^7 \yr$, and peak star formation rates $\sim
200 h^{-1} \Msol\yr^{-1}$. These are similar properties to those
inferred observationally for the ULIGS \citep[\eg][]{sm96}.

\subsection{Properties of UV-bright starbursts}
\label{sec:burst}

A large amount of work has been done on samples of UV-bright starbursts
selected from the catalogue of UV spectra of star-forming galaxies of
\citet{kinney93}. Various correlations have been found, for instance between
the bolometric luminosity, the UV/IR ratio, the slope of the UV continuum and
the metallicity \citep[\eg][]{meurer95,heck98}. In this section, we compare the
properties of our model starbursts with some of this observational data.

The observational sample that we use for our comparison is that of
\citet{heck98}, who selected 45 starburst and star-forming galaxies
from the original atlas of \citeauthor{kinney93}. The criteria for a
starburst galaxy to appear in the \citeauthor{kinney93} catalogue are
(a) that it has been previously classified as a starburst based on
optical data, usually meaning that it has a compact optical morphology
and strong optical emission lines (but no AGN activity)
\citep[\eg][]{balzano83}; and (b) that it has been observed by IUE and
has a high enough surface brightness within the IUE aperture to
produce a reasonable quality UV spectrum. The catalogue is not in any
sense statistically complete. The starburst activity in these galaxies
is generally confined to the central regions. (The galaxies have
mostly been selected so that the starburst activity fits within the
IUE aperture, $20"\times 10"$, while the optical diameters of the
underlying galaxies are typically a few arcminutes.)

For the galaxies in their sample, \citet{heck98} measured a UV luminosity
$L_{UV} \equiv \lambda L_{\lambda}(1900\AA)$ and mean continuum slope $\beta$
between 1250 and 1850 \AA\ (defined by $L_{\lambda} \propto \lambda^\beta$)
from IUE spectra, and a far-IR luminosity $L_{FIR}$ from IRAS measurements.
\citeauthor{heck98} use the definition of $L_{FIR}$ from \citet{helou88}, which
can be expressed as in terms of the luminosities in the 60 and 100 $\mum$ IRAS
bands as
\begin{equation}
L_{FIR} = 0.65 \nu L_{\nu}(60) + 0.42 \nu L_{\nu}(100)
\label{eq:LFIR}
\end{equation}
$L_{FIR}$ provides an estimate of the $40-120\mum$ luminosity. The quantity
$L_{FIR}+L_{UV}$ is similar to the bolometric luminosity in the case of
starbursts, where most of the radiation is emitted in either the UV or the FIR.
The galaxies in the \citeauthor{heck98} sample cover the range $L_{FIR}+L_{UV}
\sim 10^8-10^{11} \Lsol$.

\begin{figure}[!htb] 
\epsscale{1.0}
\plotone{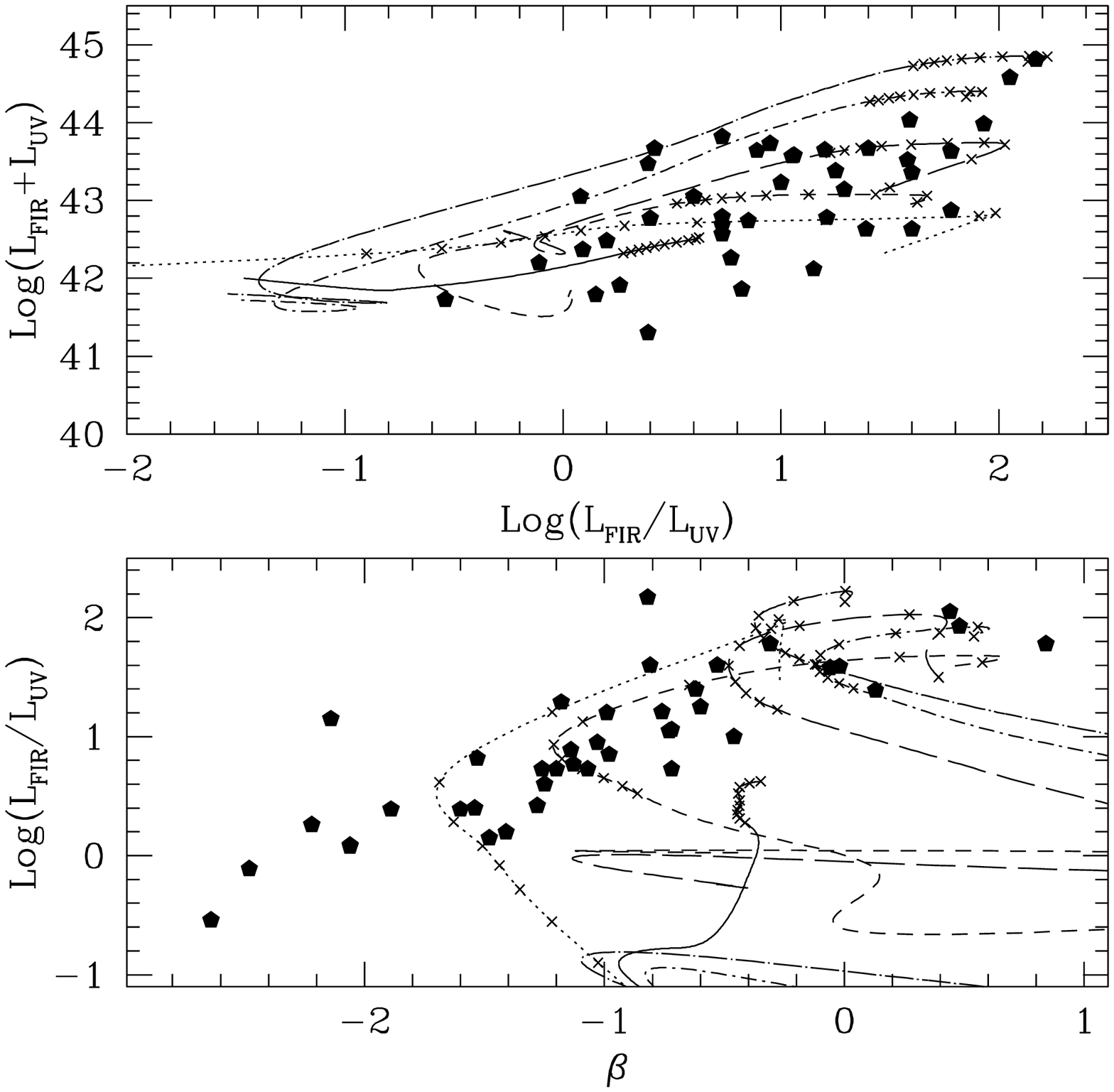}
\caption{Upper panel: $L_{FIR}+L_{UV}$ (roughly corresponding to the
bolometric luminosity) in $h^{-2}\ergs$ versus the ratio of the IR to
the UV luminosity $L_{FIR}/L_{UV}$. Lower panel: the relation between
$L_{FIR}/L_{UV}$ and the slope $\beta$ of the UV continuum. In each
panel, the filled symbols show observed local starbursts from
\protect\citet{heck98}, while the lines show the evolutionary tracks
of five different model starbursts. The bursts start towards the upper
right part of each panel, and evolve towards the lower left (top
panel) or lower right (bottom panel). The crosses mark the age since
the start of the burst in steps of 5 Myr.}
\label{fig:path}
\end{figure}

\begin{figure}[!htb] 
\epsscale{1.0}
\plotone{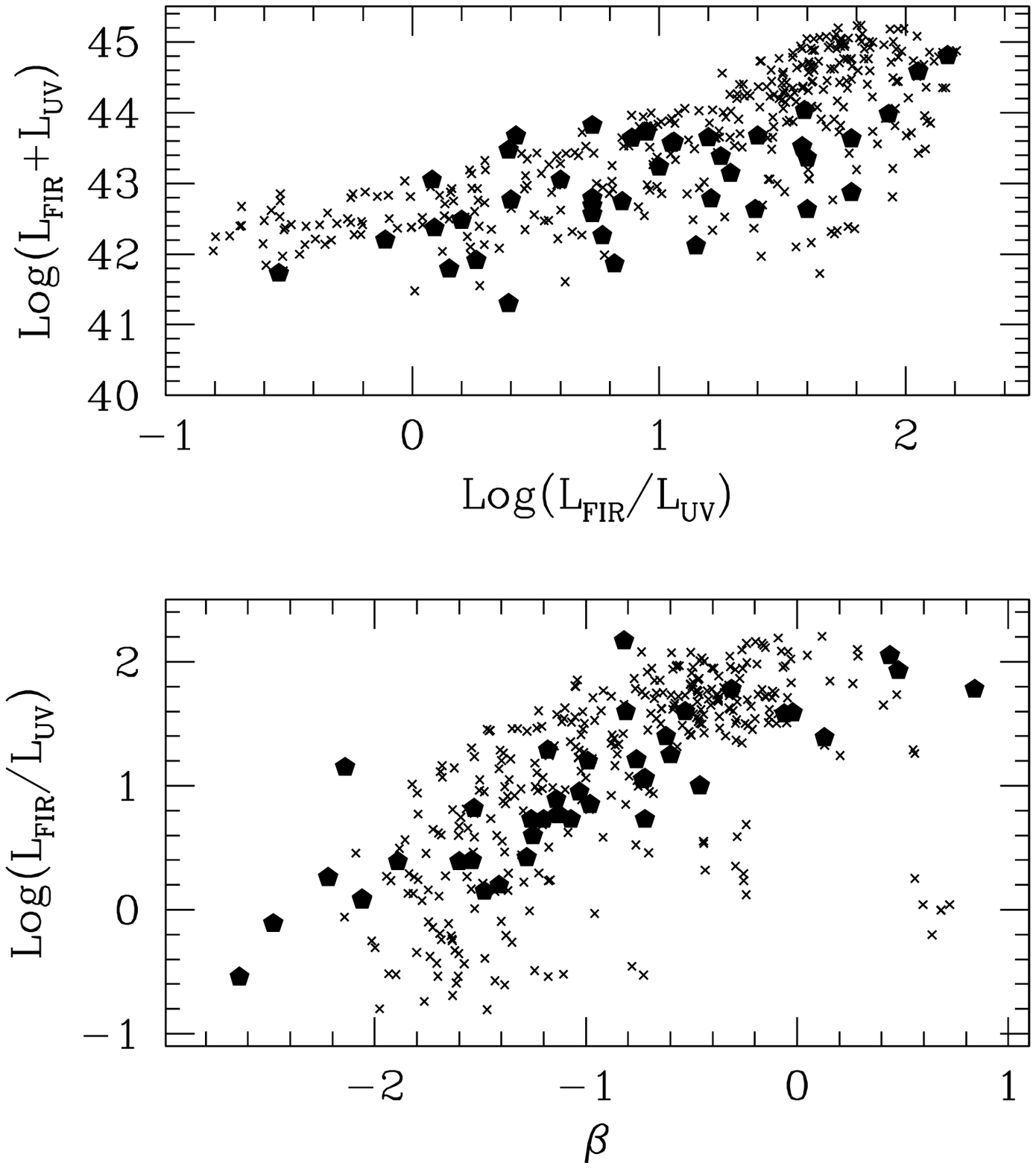}
\caption{Same as in Fig.~\ref{fig:path} but for a random sample of
model bursts (crosses), as described in the text.  The filled
pentagons are the same observational data from
\protect\citet{heck98}.}
\label{fig:bol}
\end{figure}

The evolutionary tracks of a selection of model starbursts, with burst
masses covering the range $\Mburst\sim 10^7 - 10^{10}\Msol$, in
$L_{FIR}+L_{UV}$, $L_{FIR}/L_{UV}$ and $\beta$ are shown in
Figure~\ref{fig:path}, together with observational data for the
\citeauthor{heck98} sample. We have calculated these quantities from
the model SEDs to match the way they are calculated from the
observational data. The bolometric luminosities of the model bursts,
as measured by $L_{FIR}+L_{UV}$, peak soon after the start of the
burst, following which they evolve towards smaller values. At the same
time, the amount of dust reprocessing of the radiation, as measured by
$L_{FIR}/L_{UV}$, also decreases.  This results from two effects: the
escape of young stars from the dense molecular clouds, and the
decrease in the optical depth of the diffuse dust component as the gas
in the burst is consumed. The UV slope $\beta$ initially evolves
towards more negative values, \ie {\em bluer}, as the net dust opacity
falls. However, as the rate of formation of new stars drops and the
dominant stellar population becomes older, the {\em intrinsic}
unabsorbed stellar spectrum becomes redder, so the evolution in
$\beta$ reverses, the models becoming redder with time even though the
dust attenuation is falling. This happens after $20-30 \Myr$,
controlled mainly by the stellar evolution timescale. As long as the
evolution in $\beta$ is dominated by the declining dust opacity, the
models stay close to the locus of observed points in the
$L_{FIR}+L_{UV}$ vs $L_{FIR}/L_{UV}$ and $L_{FIR}/L_{UV}$ vs $\beta$
panels, but when the intrinsic stellar spectrum starts to redden with
age, the models move away from the observed locus. This is not in
itself in contradiction with observations, since there are selection
effects in the observational sample, as we discuss below.

The burst evolution involves the interplay between two timescales, the lifetime
of massive stars, $\sim 10^7 \yr$, and the exponential decay time $\tau_e$ of
the star formation rate and gas mass in the burst. The latter varies with burst
mass, being larger than the stellar evolution timescale for large bursts, and
comparable for small bursts, as shown in Fig~\ref{fig:burstprops}. The model
starbursts begin their evolution with a large infrared excess and a flat UV
slope (upper and lower panel of Fig.~\ref{fig:path}). Fainter bursts, which
have lower gas column densities and are on average also more metal poor,
quickly exhaust their gas content and evolve toward a low infrared excess and a
negative UV slope, along the locus defined by observations (lower panel of
Fig.\ref{fig:path}). Conversely, brighter bursts, having larger gas column
densities and higher metallicities, remain highly enshrouded by dust until,
after a few tens of Myr, the dominant stellar population has become
intrinsically old. Their UV continuum slopes always remain flat, at the
beginning because of reddening and at later times because of age.

In summary, the model bursts lie close to the region occupied by observed
bursts in Fig.~\ref{fig:path} as long as the stellar population is young, in
the sense of the UV light being dominated by very massive stars. The position
of bursts along the observational locus is then determined mostly by the net
dust opacity in the UV, in agreement with the interpretation of
\citet{meurer95} and \citet{heck98}. This in turn depends both on the initial
gas mass, radius and metallicity of the burst, and on its evolutionary stage.

A detailed comparison with the \citeauthor{heck98} observations would require
us to construct a mock catalogue of model starbursts obeying the same selection
criteria as the observed sample. Unfortunately, the observational selection
criteria are rather ill-defined. In addition, one of the selection criteria is
the presence of strong $HII$ region emission lines, and the GRASIL code at
present does not calculate these emission line properties.  Instead, we simply
select starbursts with ages since the start of the burst less than $t_{max}$,
to account roughly for the effect that as soon as most of the massive stars
have evolved away, the galaxy will no longer produce strong emission lines, and
so no longer be classified as a burst in the observational sample.
Fig.~\ref{fig:bol} shows the resulting distribution of points for the choice
$t_{max}=50 \Myr$. The model starbursts are seen to follow similar relations to
the observational sample. The results do not depend sensitively on the choice
of t$_{max}$.

Several parameters may in principle affect the the spectral properties of a
model starburst galaxy and therefore the location of our models in the above
plots, but the most critical are the ratio between the star formation timescale
and the dynamical time ($\epsburst^{-1} = \taustburst/\taubulge$), and the
escape time ($\tesc$) for newly born stars to escape from their parental
clouds. The former affects the bolometric luminosity, which is almost directly
proportional to the star formation rate, and thus inversely proportional to the
star formation timescale. The latter affects the fraction of light absorbed
inside clouds, and so may affect both the slope of the UV spectrum and the
ratio between the IR and UV luminosities.  The distribution of model points in
Fig.~\ref{fig:bol} can therefore be used to constrain the values of
$\epsburst$ and $\tesc$. However, we found that changes in either of these
parameters by a factor $\sim 2$ either way would only slightly worsen the match
with observations.

\begin{figure}[!htb] 
\epsscale{1.1} 
\plotone{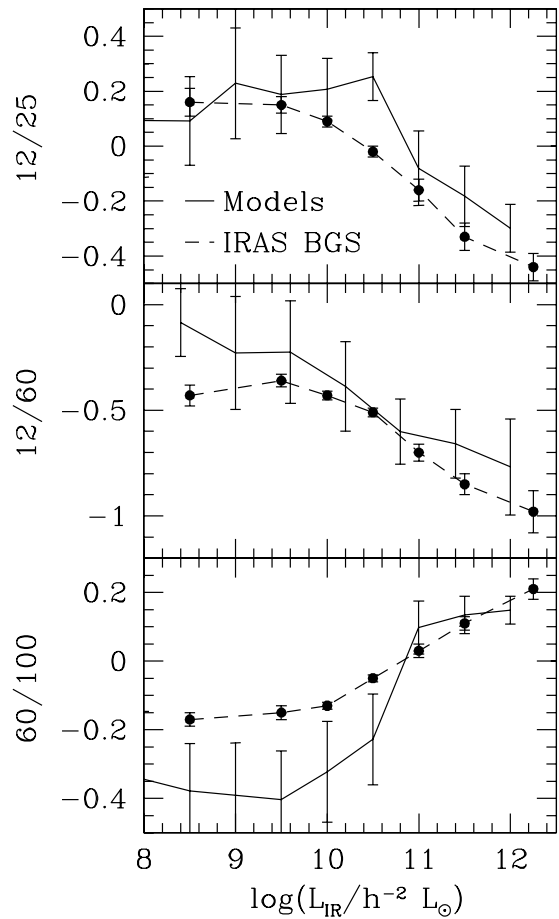}
\caption{Infra-red colours versus total infra-red luminosity. The
three panels show the 12/25, 12/60 and  60/100 $\mu$m colours
respectively, defined as the log of the ratio of $\nu L_\nu$ in each band. 
In each panel, the solid line with error bars shows the mean and
dispersion from the models, while the dashed line with error bars
shows the mean and error on the mean measured by \protect\citet{sn91}
from the IRAS Bright Galaxy Survey.}
\label{fig:soif}
\end{figure}

\subsection{Infrared colours}

We now consider the infra-red and sub-mm colours of starbursts and normal
galaxies. Figure~\ref{fig:soif} shows the dependence of the mean IRAS colours
on infra-red luminosity. This plot includes all model galaxies, both normal and
starbursts. Their IRAS band luminosities are calculated by convolving the SEDs
with the IRAS response functions. In calculating the mean colours, the models
are weighted by their number density and by a factor $L_{\nu}^{3/2}(60\mum)$ to
account approximately for the volume within which a galaxy would be visible in
a $60\mum$ flux-limited sample. $L_{IR}$ is the standard estimate of the total
$8-1000\mum$ IR luminosity from the luminosities in the four IRAS bandpasses
\citep{sm96}:
\begin{eqnarray}
L_{IR} &=& 0.97 \nu L_\nu(12) + 0.77 \nu L_\nu(25) \nonumber \\
&+&  0.93 \nu L_\nu(60) + 0.60 \nu L_\nu(100).  
\label{eq:lir}
\end{eqnarray}
The model predictions are compared with the observed mean colours calculated by
\citet{sn91} from the IRAS bright galaxy sample (IRAS BGS, \citealt{bgs}),
which is a complete sample flux-limited at $60\mum$. Models and observations,
in particular the 12/60 $\mum$ colour, show in general similar trends.

\begin{figure}[!htb] 
\epsscale{1.0}
\plotone{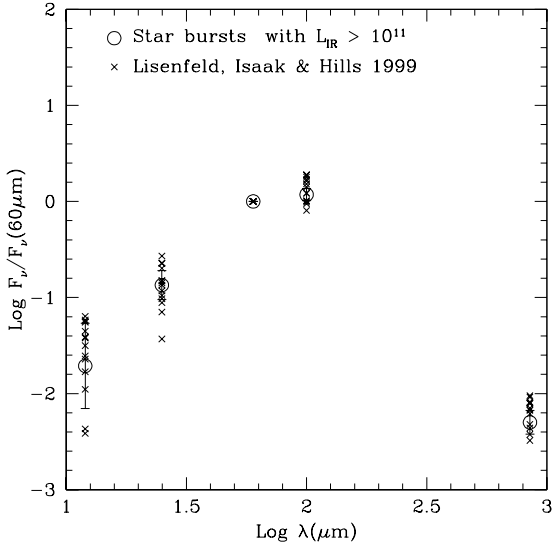}
\caption{Average SED of model starbursts with $L_{IR}\geq
10^{11}h^{-2}\Lsol$ (empty circles) compared to the observational data
(crosses) for 14 luminous starbursts from \protect\citet{lis99}. (We
exclude galaxies with only upper limits at $850\mum$.) All SEDs are
normalized to the 60$\mum$ luminosity. The error bars show the
dispersion in the models.}
\label{fig:scub}
\end{figure}

Figure~\ref{fig:scub} shows the quite good agreement between the
average IR and sub-mm SED of model starburst galaxies, and the
observed SEDs of luminous infra-red galaxies from the sample of
\citet{lis99}. The limit $L_{IR}\geq 10^{11}h^{-2}\Lsol$ for the
models has been chosen to approximately reproduce the selection for
the \citeauthor{lis99} sample.  Note that the dust opacity in our
models decreases as $\lambda^{-b}$ with $b\approx2$ for
$100\lsim\lambda\lsim 1000\mum$, while \citeauthor{lis99}, by fitting
optically thin single temperature models to the data at $\lambda\geq
60\mum$, derived $b$ values in the range 1.5--2. Our models
demonstrate that the shallower slopes can instead be explained by the
distribution of dust temperatures within each galaxy.

\subsection{Extinction in starburst galaxies}
\label{sect:ext}

An important problem in the study of star-forming galaxies is to
determine the amount of attenuation of starlight by dust, especially
in the UV. This bears directly on the determination of star formation
rates in galaxies from their UV luminosities. For our own and a few
nearby galaxies, the {\em extinction} law of the dust can be measured
directly from observations of background stars, where the dust acts as
a foreground screen. The differences found between the shapes of the
extinction curves of the Galaxy, the Large Magellanic Cloud and the
Small Magellanic Cloud below $\lambda\leq$2600\AA\
\citep[\eg][]{fitz89} are often ascribed to the different
metallicities in these systems, covering the range $Z\sim 0.1-1\
\Zsol$. Recently, \citet{cks94} \citep[see also][]{calz97,calz99} have
analyzed the dust extinction in starburst galaxies.  In this case, the
measurement of the extinction is more complicated, since one measures
the integrated light of the whole system, where stars and dust are
mixed in a complex way. From the optical and UV spectra of a sample of
UV-bright starbursts, \citeauthor{cks94} derive an average {\em
attenuation law} characterized by a shallower far-UV slope than that
of the Milky Way extinction law, and by the absence of the 2175 \AA\
feature. This is at first sight quite surprising, because the
metallicities of these galaxies are mostly similar to that of the
Milky Way, and so they might be expected to have similar dust
properties. The question is then to what degree the differences
between the starburst attenuation law and the Milky Way extinction law
are due to the geometry of the stars and dust, and to what degree they
can only be explained by differences in dust properties.

\begin{figure}[!htb] 
\epsscale{1.0}
\plotone{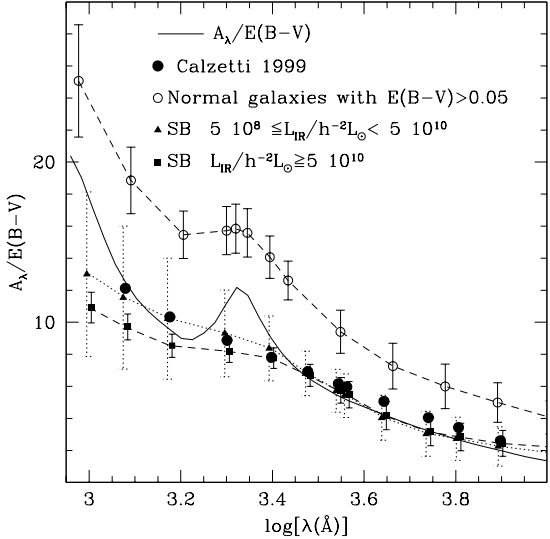}
\caption{The average dust attenuation curves for starlight in
different classes of galaxies (normal and starburst, SB) in the model
compared with the average Milky Way extinction law (solid line) and
with the Calzetti ``attenuation law'' (filled circles,
\citet{calz99a}, with $R'_V=4.05$). The attenuation curves are all
normalized to the net reddening of the stellar population,
$E(B-V)$. The error bars show the dispersion of the models around the
mean attenuation curve.}
\label{fig:calz}
\end{figure}

Figure~\ref{fig:calz} compares the average attenuation curves for
galaxies from our model with the empirical ``attenuation law''
obtained for starbursts by \citet{calz99a}. The attenuation
$A_{\lambda}$ for the models is defined as the difference in
magnitudes of the stellar luminosity $L_{\lambda}$ of a galaxy with
and without dust, and is normalized to the colour excess $E(B-V) =
A_B-A_V$ of the stars to give an attenuation ``law'' $k(\lambda)=
A_{\lambda}/E(B-V)$, equivalent to the definition of
\citeauthor{cks94}. As described in \S\ref{sec:grasil_dust}, the dust
properties we adopt imply an extinction law characterized by a
distinct 2175 \AA\ feature produced by graphite grains, and well
matching the average Milky Way extinction curve. The model extinction
law (solid line in Fig.~\ref{fig:calz}) is the attenuation law that
would be measured if all the dust were in a foreground screen in front
of the stars and no scattered light reached the observer. This
geometry is clearly not realistic as applied to the integrated light
from galaxies. In our models, we have instead a complex and wavelength
dependent geometry, where the UV emitting stars are heavily embedded
inside molecular clouds, while the older stars, mainly emitting in the
optical and near infrared, are well mixed with the diffuse
interstellar medium.

Figure~\ref{fig:calz} shows average attenuation curves for 3 classes
of model galaxies: (a) normal galaxies with $E(B-V)>0.05$; (b)
starbursts with $5\times10^8 < L_{IR} < 5\times10^{10} h^{-2} \Lsol$;
and (c) starbursts with $L_{IR} > 5\times10^{10} h^{-2} \Lsol$. The
starburst models are all chosen to have ages $<50 \Myr$ since the
start of the burst, as discussed in \S\ref{sec:burst}. Sample~(b)
corresponds roughly to the galaxies for which \citeauthor{cks94}
measured their attenuation law. The model attenuation law depends
significantly on the sample, but all 3 classes show a weak or
completely absent $2175\AA$ feature. In particular, the predicted
attenuation curve for the lower luminosity starbursts is remarkably
close to the empirical ``Calzetti law''. This result is an entirely
geometrical effect, and did not require us to assume for starbursts
dust properties different from those of the Galaxy.
This conclusion is contrary to that of \citet{gordon97}, who argued that the
observed shape is only produced with dust that lacks the $2175\AA$ feature in
its extinction curve. The reason is presumably that \citeauthor{gordon97} only
considered clumping of dust, not of stars, and assumed a spatial distribution
for stars independent of stellar age. Our results follow naturally from the
assumption that stars are born inside dense dust clouds and gradually escape.

\begin{figure}[!htb] 
\begin{center}
\epsscale{1.0}
\plotone{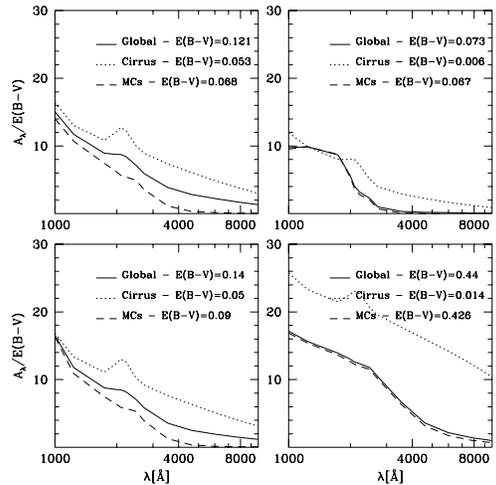}
\end{center}
\caption{Predicted attenuation curves for two normal galaxies (left
panels) and two starbursts (right panels) from the models.}
\label{fig:att}
\end{figure}

To further illustrate the importance of geometrical effects in determining the
attenuation law, we show in Fig.~\ref{fig:att} the attenuation laws of 2 normal
and 2 starburst model galaxies. The global attenuation law, and the separate
contributions from the molecular clouds and diffuse dust, are shown in each
case, normalized to the colour excess $E(B-V)$ produced by that dust component.
The global ($g$), molecular cloud (MC) and diffuse dust ($d$) contributions are
related by
\begin{equation}
(A_\lambda/E)_g=\frac{(A_\lambda/E)_{MC} E_{MC}
+(A_\lambda/E)_{d} E_{d}}{E_g}
\end{equation}
In the far-UV, including the spectral region around the $2175\AA$ feature, the
global attenuation in the models is strongly contributed, or even dominated, by
the MCs.  The shape of the attenuation curve there has little to do with the
optical properties of grains, because our MCs usually have such large optical
depths that the UV light from stars inside the clouds is completely absorbed.
The wavelength dependence of the attenuation law of the MC component instead
arises from the fact that the fraction of the light produced by very young
stars increases with decreasing wavelength, and at the same time, the fraction
of stars which are inside clouds increases with decreasing age, as given by
eqn.~(\ref{eq:tesc}). The additional attenuation arising in the cirrus
component can sometimes imprint a weak $2175 \AA$ feature, but this is not the
case for the starbursts, where the primary UV stellar light is dominated
by very young populations.


\section{Galaxy Luminosity Function}
\label{sec:lfs}

\subsection{Method}

The luminosity function of galaxies at different wavelengths is a
basic property of the galaxy population which a galaxy formation model
should explain. We calculate the galaxy luminosity function at
different wavelengths by combining the model SEDs with the weights for
the individual galaxies (Section~\ref{sec:procedure}). 
For the normal galaxy sample we have,
for the number density of galaxies per $\ln L$ at some wavelength
$\lambda$
\begin{equation}
{dn \over d\ln L_{\lambda}} = {1\over \Delta\ln L} \sum_{|\ln L_i-\ln L| <
\frac{1}{2}\Delta(\ln L)} n_i
\end{equation}
where $n_i$ is the number density for the $i$th galaxy, $L_i$ is its
luminosity at wavelength $\lambda$, the centre of the bin is at $L$
and its width is $\Delta(\ln L)$. For the burst galaxy sample, we have
to sum over the burst phase $j$ also, giving
\begin{equation}
{dn \over d\ln L_{\lambda}} = {1\over \Delta\ln L} \sum_{|\ln
L_{ij}-\ln L| <\frac{1}{2}\Delta(\ln L)} n_{ij}
\end{equation}
where $n_{ij}$ is the number density for galaxy $i$ at evolutionary
phase $j$, and $L_{ij}$ its the luminosity at that phase. 

Galaxy luminosity functions are measured in specific bands defined by a
filter+instrument response function, \eg the standard B or K bands, or the IRAS
bands. Thus we convolve the model SEDs with the appropriate response function
to calculate the luminosity $L_{\nu}$ in that band. We use absolute magnitudes
on the AB system, $M_{AB} = -2.5\log_{10}(L_{\nu}/4.345\times
10^{20}\ergs\Hz^{-1})$. The model luminosity functions have statistical
uncertainties due to the finite size of the model galaxy catalogue. We estimate
these statistical errors by bootstrap resampling of the catalogue.

\begin{figure*}[!htb] 
\epsscale{1.5} 
\plotone{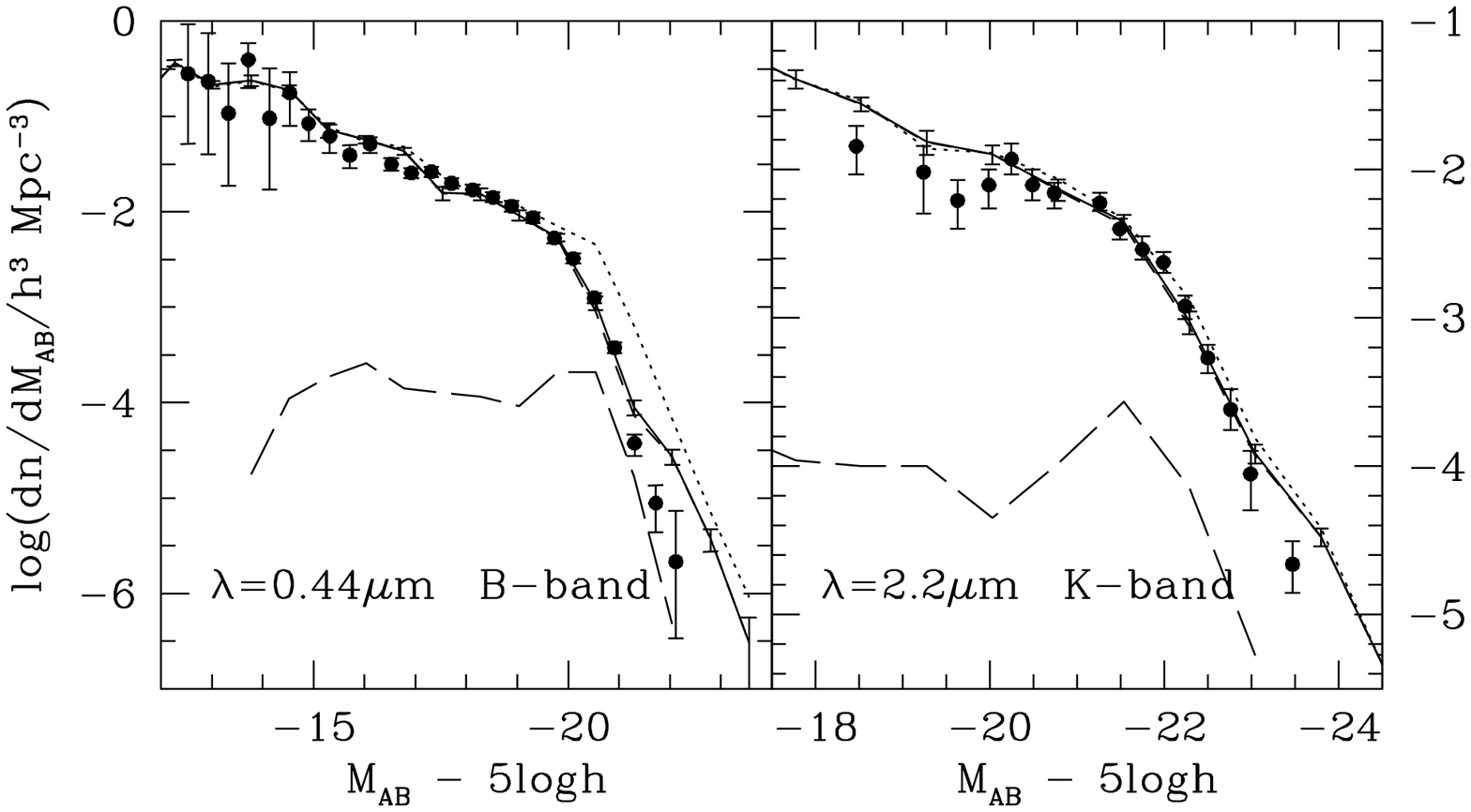}
\caption{Galaxy luminosity functions in the optical and near-IR. The left
panel shows the luminosity function in the B-band ($0.44 \mu m$),
compared to observational data from \protect\citet{zucca97}. The right
panel shows the luminosity function in the K-band ($2.2 \mu m$),
compared to observational data from \citet{gardner97}. In both panels,
the solid line shows the total luminosity function including the
effects of dust, while the long-dashed and short-dashed lines show the
contributions to this from galaxies with and without recent bursts,
and the dotted line shows the luminosity function without dust. The
error bars on the model luminosity functions are the statistical
errors resulting from the finite sample size, calculated using
bootstrap resampling. The observational data have been converted to AB
magnitudes assuming $B=B_J+0.2$ (to convert from measured to standard
magnitudes) and $B_{AB}=B-0.12$, and $K_{AB}=K+1.87$ respectively.}
\label{fig:lfbk}
\end{figure*}

\subsection{Optical and Near Infra-Red}

At optical and near-IR wavelengths the emission is mostly from older stars and
the effects of dust obscuration are generally modest. Figure~\ref{fig:lfbk}
shows the local luminosity function in the B-band ($\lambda=0.44\mum$),
compared to the observed luminosity function measured from the ESP redshift
survey by \citet{zucca97}. The predicted luminosity function agrees well with
the observed one, except at the highest luminosities. Extinction by dust makes
galaxies around $0.6 \mag$ fainter on average, for bright ($L\gsim L_\star$)
galaxies. Galaxies which have had recent bursts (\ie in the last $1/20$ of the
age of the universe,  0.7 Gyr) do not dominate the luminosity function at any
luminosity, when the effects of dust are included.

As described in \citet{cole99}, the B-band luminosity function is used as one
of the primary observational constraints for setting the parameters in the
GALFORM model, in particular, the parameters $\alphahot$ and $\Vhot$
controlling feedback, and the parameter $\Upsilon$ which sets the fraction of
brown dwarfs in the IMF. The good agreement with the observed B-band luminosity
function is therefore not a surprise, but it was not guaranteed, since the
stellar population and dust models used in \citet{cole99} are not identical to
those used here. Cole \etal used the stellar population models of \citet{bc99},
and calculated the effects of dust using the models of \citet{ferrara99}. The
stellar population model in GRASIL is based on similar stellar evolution tracks
and spectra, but the treatment of dust extinction is significantly different.
The Ferrara \etal models assume that stars and dust are smoothly distributed,
while in GRASIL a fraction of the dust is in clouds, and young stars are
confined to these clouds. The B-band luminosity functions, both with and
without dust, calculated by GALFORM using the \citet{bc99} and
\citet{ferrara99} models agree very well with those computed using the GRASIL
stellar population+dust model, demonstrating the consistency of the procedure
of using the galaxy formation parameters derived in Cole \etal in combination
with the GRASIL model. The effects of dust computed using the two models are
quite similar in the B-band, in spite of the differences in the star and dust
geometry. This is because most of the B-band light is produced by stars which
are old enough to have escaped from the clouds in which they formed, so in
GRASIL the attenuation is due mostly to the diffuse component of the dust,
which is modelled in a similar way to that in the Ferrara \etal models.

Figure~\ref{fig:lfbk} also shows the model and observed luminosity
functions in the K-band. In this case, the effects of dust are very
small, so the comparison is essentially independent of assumptions
about dust. Again, the model agrees well with observations over most
of the luminosity range, as was also found by \citet{cole99}. The
contribution of galaxies with recent bursts is very small at all
luminosities.

\begin{figure*}[!htb] 
\epsscale{1.5} 
\plotone{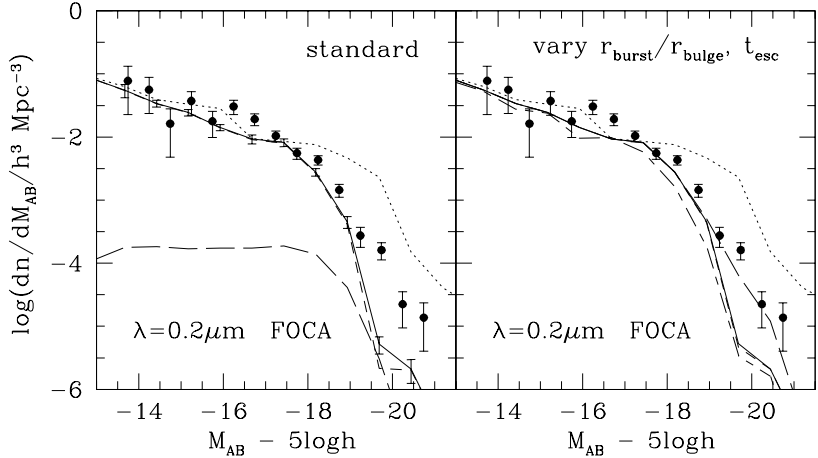}
\caption{Far-UV ($0.2\mum$) luminosity function. (a) The left panel
shows the results for our standard parameters. The solid line shows
the model prediction for the LF at $z=0$ including dust, while the
long-dashed and short-dashed lines show the contributions to this from
galaxies with and without recent bursts. The dotted line shows the
model luminosity function obtained if absorption by dust is
neglected. The symbols with error bars show the observed $0.2\mum$
luminosity function (without any dust correction) measured from a
UV-selected redshift survey by \protect\citet{sullivan99}. FOCA
magnitudes are converted to the AB system, assuming
$m_{AB}=m_{2000}+2.26$. (b) The right panel shows the effects of
varying $\tesc$ and $\rburst$.  The solid line is for our standard
model ($\tesc= 2$ and $10 \Myr$ in normal galaxies and bursts
respectively, and $\rburst/\rbulge=0.1$), including dust. The short-dashed
line shows the effect of increasing $\tesc(burst)$ to $30\Myr$, and
the short-dash-long-dash line the effect of increasing $\tesc(normal)$
to $5\Myr$. The long-dashed line shows the effect of increasing
$\rburst/\rbulge$ to 0.5. The dotted line shows the LF without dust,
which is the same in each of these cases. The observational data are
as in the left panel.}
\label{fig:lf0.2}
\end{figure*}

\subsection{Far Ultra-Violet}

In Figure~\ref{fig:lf0.2} we compare the predicted luminosity function in the
far-UV ($\lambda=0.2\mum$) with that measured by \citet{sullivan99} from a
UV-selected redshift survey, based on FOCA instrument fluxes. This comparison
has not previously been made for any semi-analytical galaxy formation models.
The effect of dust are much larger than in the optical, as one would expect. In
this case, the effects of the more realistic geometry for the stars and dust
assumed by GRASIL compared to the Ferrara \etal models (clumpy rather than
smooth distributions for the stars and dust) are significant. The
stars that produce most of the UV light spend a large fraction of their
lifetimes in the molecular clouds where they form, so the mean extinction is
larger than in the case of a smoothly distributed dust component with the same
total dust mass. Bursting and non-bursting galaxies contribute roughly equally
at the highest luminosities. This result is however sensitive to the details of
how bursts are modelled, since this determines what small fraction of the UV
light escapes from currently or recently bursting galaxies. When we compare our
model LF including extinction with the directly observed LF, uncorrected for
extinction, we find reasonable agreement at lower luminosities, but at high
luminosities, the model LF is somewhat lower than the observed one. This might
be partly an effect of evolution in the observational sample, which covers a
significant redshift range ($z\lsim 0.5$), but it might also be that the UV
extinction is over-estimated in the model.

Figure~\ref{fig:lf0.2} shows also the effect of changing the burst radius
$\rburst$ and the timescale $\tesc$ for stars to escape from clouds.
Increasing $\rburst/\rbulge$ from 0.1 to 0.5 reduces the optical depth in the
diffuse component during bursts, allowing more of the UV light from bursts to
escape, and increasing the LF at the highest luminosities. Increasing $\tesc$
in bursts from $10 \Myr$ to $30 \Myr$ has negligible effect on the
total UV LF. Increasing $\tesc$ in normal galaxies from $2\Myr$ to
$5\Myr$ slightly lowers the amplitude of the luminosity function at
the bright end.

\begin{figure*}[!htb] 
\epsscale{1.5} 
\plotone{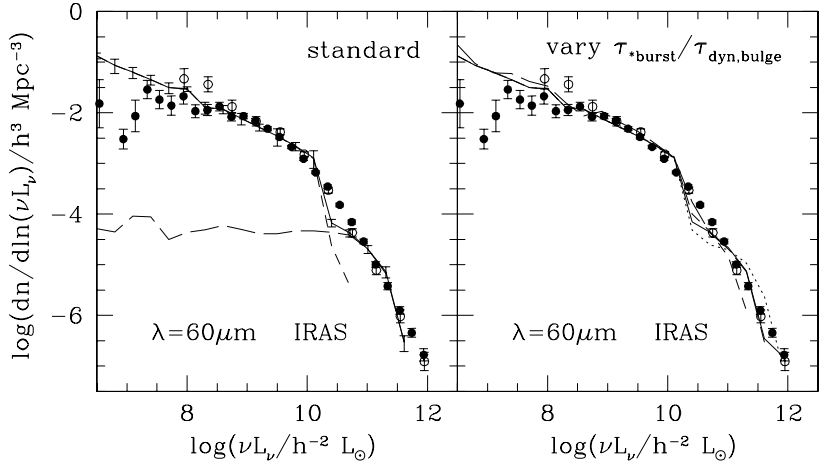}
\caption{The $60\mum$ far-IR luminosity function. (a) The left panel
shows the results for our standard parameters. The solid line shows
the total model luminosity function, while the long-dashed and short-dashed
lines show the contributions to this from galaxies with and without
recent bursts. The points with error bars show the observed $60\mu m$
luminosity functions measured from galaxies observed by IRAS, by
\citet{saunders90} (filled circles) and \citet{sn91} (open
circles). (b) The right panel shows the effects of
varying the burst or escape timescales: $\epsburst^{-1}\equiv
\taustburst/\taubulge =$ 1 (dotted), 2 (solid), 4 (dashed). 
The long dashed curve 
has $\epsburst^{-1}=2$ (our standard value) but $\tesc = 2$
Myr instead of 5 Myr.}
\label{fig:lf60}
\end{figure*}

\subsection{Mid and Far Infra-Red}
\label{sec:irlf}

In the mid- and far-infrared, the luminosity of galaxies is dominated by
re-emission from dust. Using the GRASIL code, we can now directly predict the
far-IR luminosities of galaxies from our galaxy formation model, and compare
with observations. The luminosity functions of galaxies at 12, 25, 60 and
100$\mum$ have been measured using IRAS data. The best determination is at
$60\mum$, where IRAS was most sensitive. Figure~\ref{fig:lf60} shows that the
predicted luminosity function agrees extremely well with that observed by
\citet{saunders90} and \citet{sn91}, except at very low luminosities, where the
measured LF is fairly uncertain. Above $\nu L_\nu(60) \gsim 3\times 10^{10}
h^{-2} \Lsol$, the model LF is dominated by galaxies undergoing bursts
triggered by mergers. This is in agreement with observations of ultra-luminous
IRAS galaxies, which are all identified as recent mergers based on their
optical morphology \citep[\eg][]{sm96}.

The right panel of Figure~\ref{fig:lf60} shows the effect on the
$60\mum$ LF of varying the parameter $\epsburst$, which relates the
star formation timescale in bursts to the dynamical time of the bulge
(equation~\ref{eq:tauburst}).  Unlike the other parameters in the
GALFORM model, \citet{cole99} did not try to choose a best-fit value,
because the observational data in the optical and near-IR that they
compared with were not sensitive to its value. (The Cole \etal results
were calculated assuming $\taustburst=0$.) However, the far-IR LF is
sensitive to this and thus constrains the burst timescale for the most
luminous galaxies. Figure~\ref{fig:lf60} shows predictions for
$\epsburst= $ 1, 0.5, 0.25, corresponding to
$\taustburst/\taubulge=1,2,4$ respectively. Increasing $\epsburst$
means bursts are more luminous, but last for a shorter time, and so
have a lower number density. This trend is seen at the high-luminosity
end of the $60\mum$ LF, which is dominated by bursting galaxies. A
value $\epsburst=2$ seems to fit somewhat better than higher or lower
values, so we adopt this as our standard value. Also shown in the same
panel is the somewhat better fit obtained setting $\tesc=5$ Myr. However,
as explained in \S~\ref{sec:choice}, our adopted standard value 2
Myr is favored by stellar evolution timescale argument and by the UV
LF. Increasing $\tesc$ in bursts from $10\Myr$ to $30\Myr$ has
negligible effect on the LF.

\begin{figure}[!htb] 
\epsscale{0.9}
\plotone{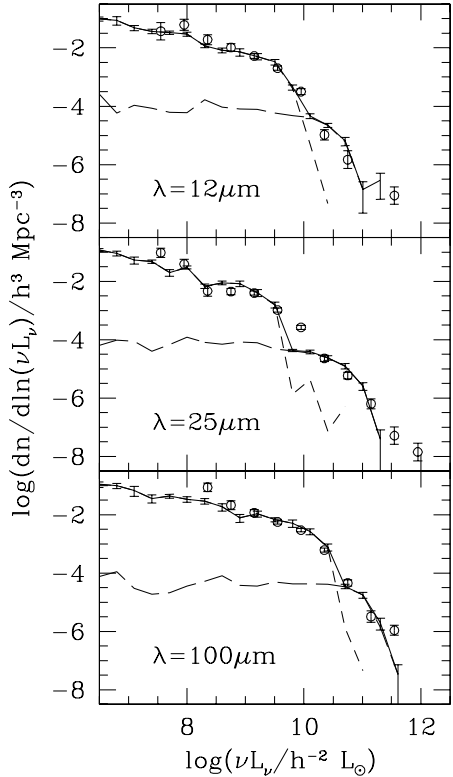}
\caption{The luminosity functions in the 12, 25 and 100$\mum$ IRAS
bands, compared to observational data from \citet{sn91}. The line
types and symbols are as in Fig.\protect\ref{fig:lf60}. }
\label{fig:ir_3lf}
\end{figure}

The luminosity functions at 12, 25 and 100 $\mum$, are compared with the
observational data from \citet{sn91} in Figure~\ref{fig:ir_3lf}. The predicted
luminosity function agrees well with the measured one in each case.


\section{Star formation rate indicators}
\label{sec:sfrcal}

Here we examine the accuracy of several SFR indicators based on
continuum UV or IR luminosities \citep[reviewed by \eg][]{kenn99}.

\begin{figure}[!htb] 
\epsscale{1.0}
\plotone{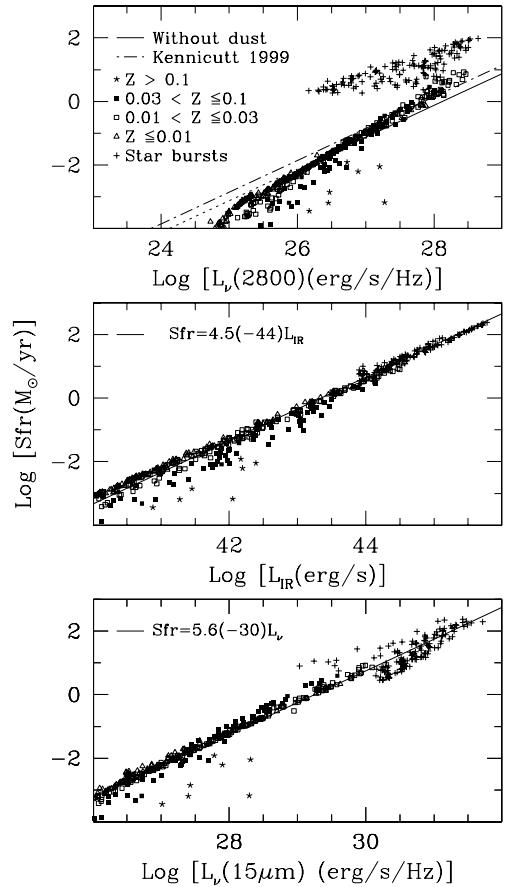}
\caption{Star formation rate versus UV and IR luminosities for
galaxies from our model. Only models corresponding to spirals and
luminous starburst galaxies are plotted. See text for details.}
\label{fig:sfr}
\end{figure}

The luminosity $L_{\nu}(2800)$ at $2800\AA$ has been extensively used to
estimate SFRs of high-redshift galaxies and to investigate the evolution of the
cosmic SFR density \citep[\eg][]{lilly96,conolly97}. In the top panel of
Fig.~\ref{fig:sfr},
we plot the SFR against $L_{\nu}(2800)$ for the model galaxies, including the
effects of extinction. Only models corresponding to spiral galaxies
(B/T $\leq$ 0.5) and luminous starburst galaxies ($L_{IR} \geq 10^{10}
h^{-2}\Lsol$) are shown. At higher luminosities and SFRs, $L_\nu(2800) \gsim
3\times10^{26} \ergs\Hz^{-1}$, the models {\em in the absence of dust} follow a
{\em linear} relation between SFR and $L_{\nu}$, with a rather small
dispersion, as would be expected if the UV luminosity is dominated by young
stars and the recent SFR has been approximately constant. This linear relation,
$SFR/(\Msol\yr^{-1}) = 8.5\times 10^{-29} L_{\nu}(2800\AA)/(\ergs\Hz^{-1})$, is
indicated by the solid line, and its extrapolation to lower luminosity is shown
by the dotted line. Dust extinction shifts points to the left of this line. At
lower luminosities, the effects of dust extinction are very small, because of
the low gas contents of the galaxies. On the other hand, because the SFRs are
so small, the $2800\AA$ light has a significant contribution from post-AGB
stars and old metal poor populations, and this causes the locus of points to
bend to the right of the linear relation. The galaxies with very high
metallicities ($Z>0.1$) have very low gas fractions.

The figure also shows as a dashed line the linear relation between
$L_\nu(2800)$ and SFR obtained by \citet{kenn99}, using stellar
population models for a Salpeter IMF, and assuming a constant SFR for
the last $10^8\yr$. The $SFR/L_{\nu}$ ratio in our models without dust
(solid line) is about 40\% lower than Kennicutt's value, but this
difference is entirely due to the different IMF we adopt
(equation~(\ref{eq:imf})).

Perhaps the most striking feature of this plot is that the starburst
models are offset by more than an order of magnitude from the average relation
holding for the normal spirals, because of the large UV extinctions in the
starbursts. Furthermore, their dispersion in SFR at a given luminosity is also
quite large. Thus, the 2800\AA\ luminosity with no dust correction performs
rather poorly as a quantitative SFR indicator, both for very high SFRs (because
of extinction) and for very low ones (because of the light from older stars).

The middle panel of Fig.\ref{fig:sfr} depicts the relation between the star
formation rate and L$_{IR}$ (eq. \ref{eq:lir}), the estimated $8-1000\mum$
luminosity based on the IRAS fluxes.  The solid line represents the relation
derived by \citet{kenn98} for starbursts, $SFR/(\Msol\yr^{-1}) = 4.5\times
10^{-44} L_{IR}/(\ergs)$, by assuming that the bolometric output in a
continuous burst of age between 10--100 Myr is completely reprocessed by dust,
again for a Salpeter IMF. The Kennicutt relation is seen to fit our model
galaxies quite well at all luminosities (normal spirals as well as starbursts),
even though we assume a different IMF from Kennicutt.

The luminosity in the ISO $15\mum$ band has also been proposed as an
approximate SFR indicator. The bottom panel of Fig.~\ref{fig:sfr} shows the
SFR vs $L_{\nu}(15\mum)$ for our galaxies. The line plotted is our best linear
fit to the model points, $SFR/(\Msol\yr^{-1}) = 5.6\times10^{-30}
L_{\nu}(15\mum)/(\ergs\Hz^{-1})$. Despite the important contribution from PAH
bands, the correlation of the SFR with $L_{\nu}(15\mum)$ in our models is still
fairly good.


\section{Summary and Conclusions}
\label{sec:conc}

We have combined an {\em ab initio} model of galaxy formation
\citep[GALFORM, \S\ref{sec:galform},][]{cole99}, with an {\em ab
initio} model for stellar emission and dust emission and absorption in
galaxies \citep[GRASIL, \S\ref{sec:grasil},][]{silva98}. Both models
are state-of-the art. We are able to predict, in the context of the
cold dark matter cosmology, the luminosities and spectral energy
distributions from the UV to the sub-mm for the whole galaxy
population, and how these change with cosmic epoch. Here we focused on
a wide range of spectrophotometric properties of present-day galaxies,
from the UV to the sub-mm, for a flat low-density cosmology
($\Omega_0=0.3$, $\Lambda_0=0.7$) with a CDM spectrum of density
fluctuations. The model is remarkably successful in explaining the UV,
optical and IR spectrophotometric and extinction properties of
galaxies in the local universe. Future papers will investigate galaxy
evolution in the UV, optical, IR and sub-mm out to high redshift.

Dust plays a dominant role in starburst galaxies, where star formation proceeds
in the central regions of a galaxy on a short timescale. We did not previously
make any detailed comparison of the properties of the starbursts predicted by
semi-analytical galaxy formation models with observational data. Here we have
shown that these properties are nicely reproduced.

Our model predicts an average dust attenuation law for starburst
galaxies that agrees remarkably well with the empirical law found by
\citet{calz99a}, although with a significant dispersion around the
mean. In particular, the 2175\AA\ bump is absent when the net
attenuation of the galaxy light is considered. This is entirely an
effect of the geometry of stars and dust in our model, and has nothing
to do with the optical properties of dust grains. Indeed, our dust
mixture would reproduce the average Milky Way extinction curve (with a
strong 2175\AA\ feature), if it were arranged in a foreground screen
geometry. The absence of this feature in the attenuation curves of
model starbursts is because in that case the dust attenuation is
dominated by molecular clouds, with the shape controlled by the
gradual escape of young stars from the clouds.

The starburst galaxies are predicted to dominate the bright end of the
luminosity function in terms of bolometric luminosity, but because of the large
extinctions in these objects, they do not make a dominant contribution to the
bright end of the luminosity function in either the UV, optical or near-IR once
dust effects are included. At these wavelengths, the luminosity function is
dominated by normal spiral and elliptical galaxies. However, the starbursts
completely dominate the bright end of the luminosity function in the mid- and
far-IR (10--100$\mum$), at total luminosities $L_{IR}\gsim 10^{11} h^{-2}
\Lsol$. Overall, the luminosity function predictions from the far-UV to the
far-IR are a remarkable success for the model, since the dust contents and
galaxy radii are predicted {\em a priori}, and the only significant adjustable
parameter in the comparison was the ratio of the burst timescale to the bulge
dynamical time, which was chosen to fit the bright end of the luminosity
function at 60$\mum$.

As expected, our models show that the UV continuum is in general a poor star
formation indicator, both because of the large variations in the amount of
extinction, and also because of the contribution from old stellar populations
in the mid-UV ($\sim 3000\AA$) in more quiescent galaxies.  The infrared
luminosity is a much more reliable SFR indicator.

The parameters values adopted here for the GALFORM model are those
chosen previously by \citet{cole99} to fit the properties of the local
galaxy population in the optical and near-infrared, apart the new ones
specifying the timescales and radii of bursts. Cole \etal also discuss
the effects of varying the 'old' GALFORM paramaters. The purpose of
this paper was to present the effects of including dust in a fixed
galaxy formation model. The treatment of dust reprocessing with GRASIL
requires some additional parameters to be set, but opens the
possibility to test semi-analytical models against the wealth of IR
and sub-mm observations already available or planned for the near
future. The adopted values for these parameters have also been guided
by the results of \citet{silva98}, who used GRASIL to reproduce
detailed SEDs of several local normal and starburst galaxies. 
Some of the effects of
variations of these parameters on SEDs, LFs and starburst properties
have been discussed in the relevant sections of
this paper. 
A more systematic parameter study will be included in future
investigations. Among the newly introduced parameters, probably the
most important ones are $\epsburst$ and $t_{\rm esc}$, describing the
timescales of bursts and the time for young stars to escape from
their parent molecular clouds. They have significant effects on (and are
constrained by) the IR and UV luminosity functions respectively.

In conclusion, this paper is a stepping stone for future work, which
will apply the same models to galaxies at high redshift. Now that
semi-analytic models can be effectively compared with infrared and
sub-millimetre observations, as well as UV and optical data, they can
be used to work towards an observationally and theoretically
consistent picture for the history of galaxy formation and star
formation in the universe.


\begin{acknowledgements} We thank the anonymous referee for a
constructive report. We acknowledge the support by the European Community
under TMR grant ERBFMRX-CT96-0086. SMC acknowledges the support of a PPARC
Advanced Fellowship, and CSF a PPARC Senior Fellowship and a Leverhulme
Research Fellowship. CGL acknowledges the support of the Danish National
Research Foundation through its establishment of the TAC, and a PPARC Visiting
Fellowship at Durham. This work was partially supported by the PPARC rolling
grant for extragalactic astronomy and cosmology at Durham.

\end{acknowledgements}


\end{document}